\documentclass[11pt,a4paper]{amsart}
\usepackage{colortbl}
\usepackage{amsmath,amsthm}
\usepackage{latexsym}
\usepackage{amsbsy}
\usepackage{amssymb}
\usepackage{verbatim}
\usepackage{graphicx}

\def\NN{\mathbb{N}}
\def\ZZ{\mathbb{Z}} 

\def\RR{\mathbb{R}}
\def\CC{\mathbb{C}}
\def\PP{\mathbb{P}}
\def\II{\mathbb{I}}
\def\EE{\mathbb{E}}

\def\L{\mathcal{L}}

\def\G{{\mathcal G}}

\newtheorem{thm}{Theorem}[section]

\newenvironment{defin}{\vspace{1mm}\noindent\textbf{Definition.}}{\vspace{1mm}}

\begin{document}

\title{Spectral methods for volatility derivatives}

\author{Claudio Albanese}
\address{Independent consultant}
\email{claudio\_albanese@hotmail.co.uk}
\author{Harry Lo}
\address{Imperial College London and Swiss Re}
\email{lo.harry@gmail.com}
\author{Aleksandar Mijatovi\'{c}}
\address{Department of Mathematics, Imperial College London}
\email{a.mijatovic@imperial.ac.uk}

\thanks{The authors would like to thank all three referees 
for lucid and constructive
comments.
AM would also like thank Jim Gatheral for useful conversations.}

\begin{abstract}
In the first quarter of
2006
Chicago Board Options Exchange (CBOE) introduced, as one of 
the listed products, options on its implied volatility index (VIX).
This created the challenge of developing 
a pricing framework that can simultaneously handle 
European options, forward-starts, options on the realized
variance and options on the VIX. 
In this paper we propose a new approach to this problem
using spectral methods. We use a regime switching  
model with jumps and local volatility defined in~\cite{FXrev} 
and calibrate it to the European
options on the S\&P 500 for a broad range of strikes and
maturities. 
The main idea of this paper is to ``lift'' (i.e. extend)
the generator of the underlying process 
to keep track of the relevant 
path information, namely the realized variance.
The lifted generator is too large a matrix to be 
diagonalized numerically. We overcome this difficulty by 
applying a new semi-analytic algorithm for block-diagonalization. 
This method enables us to evaluate numerically
the joint distribution between
the underlying stock price and the realized variance,
which in turn gives us a way of pricing consistently 
European options, general accrued variance payoffs
and forward-starting and VIX options.
\end{abstract}

\maketitle

\section{Introduction}
\label{sec:Intro}
In recent years there has been much interest 
in trading derivative products whose underlying 
is a realized variance 
of some liquid financial
instrument (e.g. S\&P 500)
over the 
life of the contract. The most popular payoff
functions\footnote{For the precise definition 
of these products see Appendices~\ref{subsec:Var} and 
\ref{subsec:Vol}.} 
are linear, leading to variance swaps, square root, yielding
volatility swaps, and the usual put and call payoffs defining 
variance swaptions.

It is clear that the plethora of possible derivatives on the
realized variance is closely related to the standard
volatility-sensitive instruments like vanilla options, which are
also exposed to other market risks, and the forward-starting options
which are almost pure vega bets and are mainly exposed to the movements
of the forward smile. Recently Chicago Board Options Exchange (CBOE)
introduced options on the volatility index\footnote{For 
a brief description of the these securities
see Appendix~\ref{subsec:VIX}.
For the definition of VIX see~\cite{VIX}.} (VIX) which
are also important predictors for the future behaviour of 
implied volatility.
The main purpose of this paper is to introduce a framework
in which all of the above financial instruments 
(i.e. the derivatives on the realized variance as well as the
instruments depending on the implied volatility)
can be priced
and hedged consistently and efficiently. 

Our central idea is very simple and can be described as follows.
We use a model for the underlying 
that includes local volatility, stochastic 
volatility (i.e. regime switching) 
and jumps and  can be calibrated
to the implied volatility surface for a wide variety of strikes 
and maturities (for the case of European options
on the S\&P 500 see Figure~\ref{fig:ivols}).
The stochastic process for the underlying 
is a continuous-time Markov chain,
as defined in~\cite{FXrev}, where it was used for modelling
the foreign exchange rates.  
The underlying process is stationary as can be inferred  
from the fact that 
the implied forward volatility smile behaves 
in a consistent way (see Figures~\ref{fig:fs_skew3m},
\ref{fig:fs_skew6m}, \ref{fig:fs_skew1y} 
and~\ref{fig:fs_skew2y}). 

There are two features of this model that make it possible to
obtain the distributions of the future behaviour of the implied volatility
and the realized variance of the underlying.
The first feature is the complete numerical solubility 
of the model. In other words spectral theory provides 
a simple and efficient algorithm (see~\cite{FXrev}, Section~3)
for obtaining a conditional
probability distribution function for the underlying 
between any given pair of times in the future. This property is sufficient
to determine completely the forward volatility 
smile and the distribution of
VIX for any maturity.
The second feature of the framework  
is that it allows us 
to describe 
the realized variance of the underlying
risky security as a continuous-time Markov
chain.
This 
makes it possible to deal 
with the realized variance by extending the generator
of the original process and thus obtaining a new 
continuous-time Markov chain which keeps track 
simultaneously of the realized variance and
the underlying forward rate.

\begin{figure}[ht]
\centering\noindent
\includegraphics[width =0.47\linewidth]{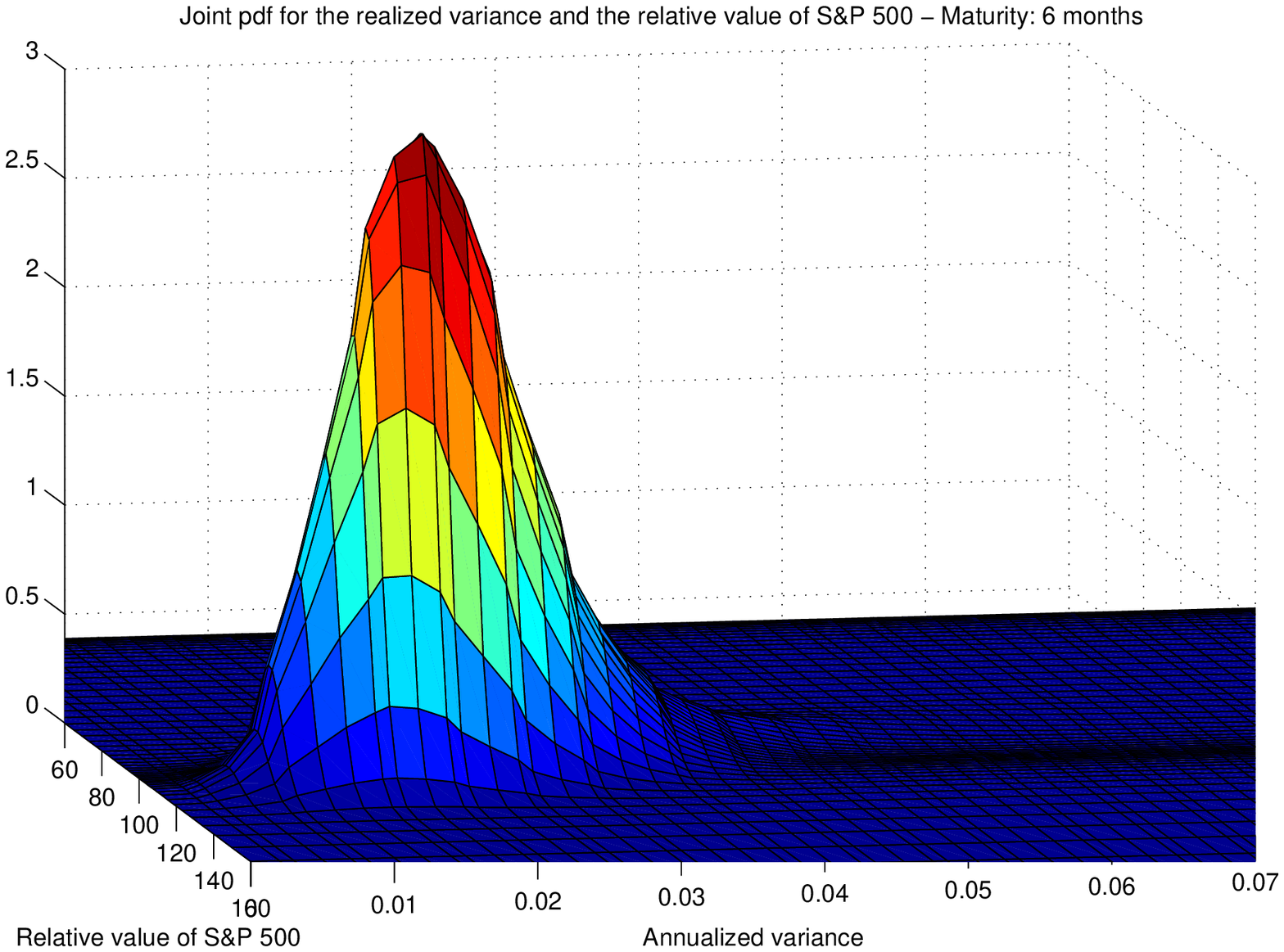}
\hfill
\includegraphics[width =0.47\linewidth]{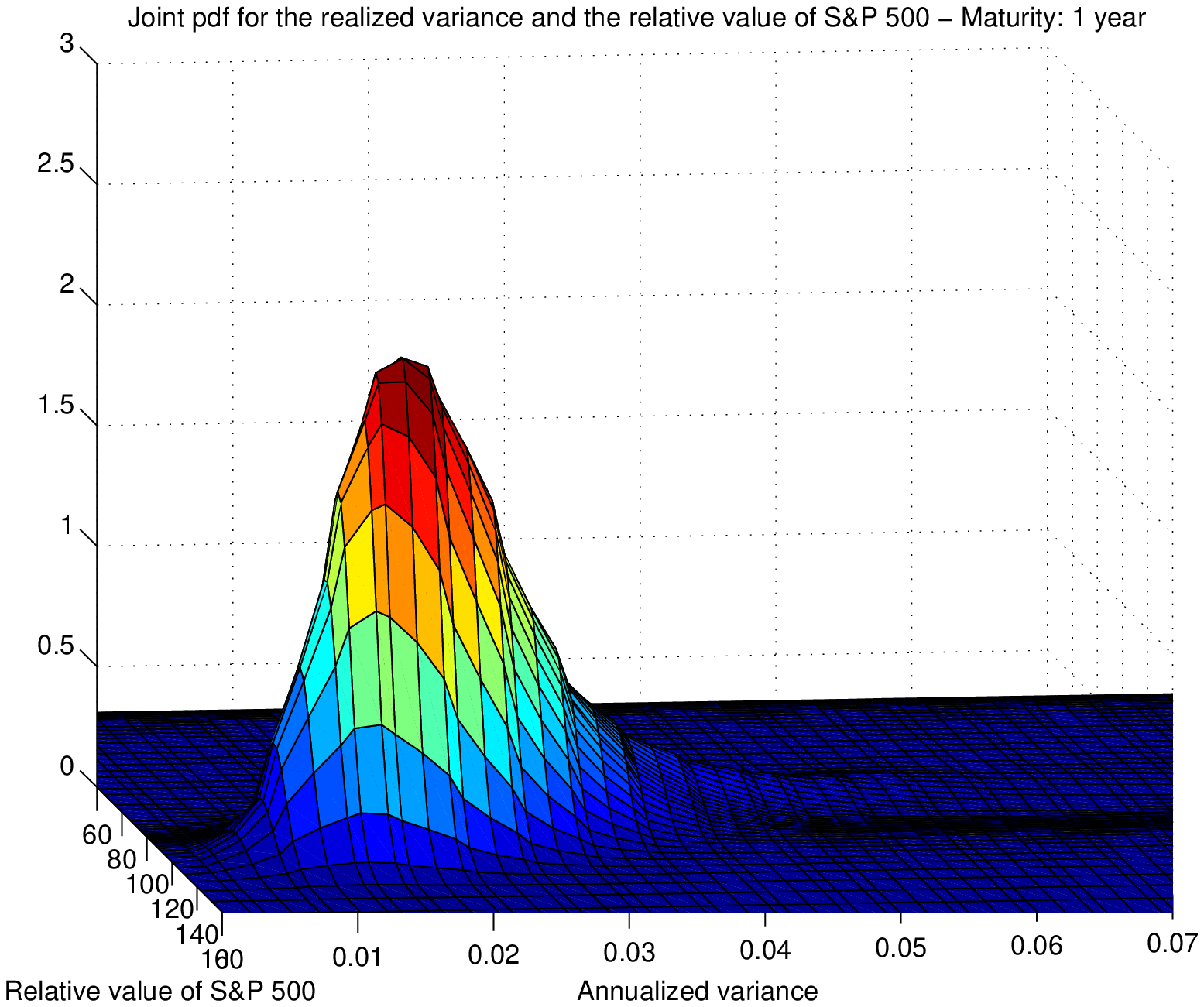}
\caption{\footnotesize{Joint probability distribution function 
                       for the annualized realized variance 
                       and the spot rate of S\&P 500 in 6 months' and 1 year's
                       time.}}
\label{fig:Joint_6m}
\end{figure}

The key idea of the paper is to assume that the
state-space of the realized variance process 
lies on a circle.
This allows us to apply a block-diagonalization
algorithm (described in detail in Appendix~\ref{sec:Bloc})
and the standard methods from spectral theory to obtain
a joint probability distribution function
for the underlying process and its realized
variance or volatility at any time in the future (e.g. for time horizons
of 6 months and 1 year 
see Figure~\ref{fig:Joint_6m}).
This joint pdf is precisely what is needed 
to price a completely general payoff that depends on 
the realized variance and on the underlying.

There are two natural and useful consequences of this 
approach. 
One is that we
do not need to specify exogenously the process for the variance 
and then try to find an arbitrage-free dynamics for the underlying,
but instead imply such a process from the observed vanilla market
via the model for the underlying
(a term-structure of the fair values of variance swaps
as implied by the vanilla market data and our model is 
shown in Figure~\ref{fig:VarSwapTerm}).
The second consequence is that this approach bypasses the use of Monte Carlo
techniques and therefore yields sharp and easily computable 
sensitivities 
of the volatility derivatives
to the market parameters. This is because the pricing
algorithm yields, as a by-product, all the necessary 
information for finding the required hedge ratios.

There is a rapidly growing interest in trading 
volatility derivatives in financial markets, mainly a consequence
of the following two factors. 
On the one hand, pure volatility instruments are used
to hedge the 
implicit vega exposure of the portfolios of market participants, thus
eliminating the need to trade frequently in the vanilla options market.
This in itself is advantageous because of the 
relatively large bid-offer spreads prevailing in that market.
On the other hand, volatility derivatives are a useful
tool for speculating on 
future volatility levels and for trading the difference 
between realized and implied volatility.

This interest is reflected in the 
vast amount of literature
devoted to volatility products. The analysis of the
realized variance is intrinsically easier than that 
of realized volatility because of the additivity of the
former. Under the hypothesis that the underlying price 
process is continuous the realized variance can be 
hedged perfectly  by a European contract with the logarithmic payoff
first studied in~\cite{Neuberger}
and a dynamic trading strategy in the underlying. 
This approach does not require an explicit specification
for the instantaneous volatility process of the underlying and 
can therefore be used within any stochastic volatility framework. 
This idea has been developed in~\cite{Carr} and~\cite{Derman1},
where the static replication strategy for the log contract, 
using calls and puts, is described.
Types of mark-to-market risk faced by a holder
of a variance swap are studied and classified 
in~\cite{Chriss}. A direct delta-hedging approach for the 
realized variance is given in~\cite{Heston}.
A shortcoming of pricing variance swaps 
without specifying a volatility model (as described
in~\cite{Carr} and~\cite{Derman}) is that this methodology
does not yield a natural method for the computation of the
sensitivities to market 
parameters (i.e. the Greeks). In~\cite{Howison} 
a diffusion model for the volatility process is specified
that allowed the authors to use PDE technology 
to price and hedge variance and volatility swaps, as well as
more general payoffs. 
Derivatives on the realized volatility can be considered
naturally as derivatives on the square root of the realized
variance. In~\cite{Brockhaus} the authors provide a volatility
convexity correction which relates the two families of derivatives.
Another approach, pioneered in~\cite{CarrLee},
develops a robust hedging strategy for
volatility derivatives which is analogous to the one
for variance swaps. 
This method works for continuous processes
only and is based on the observation 
that, under the continuity hypothesis, 
there is a simple algebraic
relationship between the Laplace transform
of the process and the Laplace transform of 
its quadratic variation. There are some technical 
difficulties in computing the relevant integrals,
which have been dealt with in~\cite{Friz}.
In~\cite{Windcliff} the authors investigate hedging 
techniques for discretely monitored 
volatility contracts that are independent
of the instantaneous volatility dynamics. 
Another model-independent 
hedging approach 
in an environment with jumps
for 
variance swaps is presented in~\cite{Schoutens}. 
A modelling framework, analogous to the HJM, for
the term-structure of variance swaps
has been proposed in~\cite{Buehler}. 
The starting point is the specification
of the function-valued 
process for the forward instantaneous variance, which 
yields an arbitrage-free dynamics for the underlying.
In order to be calibrated,
this model requires at time
zero the entire variance swap curve.


The paper is organized as follows. 
The key idea
that allows us to price general derivatives on the realized
variance is introduced in Section~\ref{sec:path}.
Section~\ref{sec:pricing} explains the pricing algorithm,
based on Theorem~\ref{thm:LiftKer}, 
for derivatives on the realized variance.
In Section~\ref{sec:calibration}
we discuss the calibration of the model to a wide range 
of strikes and maturities for options written on the 
S\&P 500.
In Section~\ref{sec:NumRes}
we carry out numerical experiments and consistency tests
on the calibrated model, including a comparison 
with a Monte Carlo method (see Subsection~\ref{subsec:Comp}).
Concluding remarks are contained in Section~\ref{sec:Conclusion}.
The numerical algorithm required to make this idea applicable
is described in Appendix~\ref{sec:Bloc}, which also contains
the proof of Theorem~\ref{thm:LiftKer}.
Appendix~\ref{sec:Vol} describes some of the volatility
contracts that can be priced within our framework
and are referred to throughout the paper.

\section{The variance process for continuous-time Markov chains}
\label{sec:path}
As mentioned in the introduction
the main idea of this paper is to describe 
the realized variance of the underlying
risky security as a continuous-time Markov
chain on a state-space which lies on a circle.
The model for the underlying 
$F_t$
will be given in 
the forward measure as a 
mixture of local and stochastic volatility 
coupled with an infinite activity jump process.
It will be defined on a 
continuous-time lattice in a largely non-parametric fashion,
as described in~\cite{FXrev}, where it was used for 
modelling the foreign exchange rate. We will briefly recall the
definition of the model in Section~\ref{sec:pricing} and 
describe its calibration to the vanilla surface on the S\&P 500 
in Section~\ref{sec:calibration}.

In the current section and the one that follows
our task is to 
build a mechanism that will make it possible 
to identify the random behaviour of the 
realized variance of the process
$F_t$.
The approach we take will only use the fact that
the process
$F_t$
is a continuous-time Markov chain
and will not depend on any other properties
of the process.
In order to do this we must return briefly to the fundamental 
theory (see~\cite{Norris}, Chapters~2 and~3, for background on
continuous-time Markov chains).

Let
$\L$
be a generator for a 
continuous-time Markov chain 
$F_t$
defined on 
a finite state space
$\Omega$.
In other words
the operator
$\L$
is given by a square 
matrix
$(\L(x,y))_{x,y\in\Omega}$
which satisfies probability conservation 
and has non-negative elements off the diagonal. 
Each element 
$\L(x,y)$
describes the first order change of the probability 
that the chain
$F_t$
jumps from level
$F(x)$
to the level
$F(y)$
in the time interval
$[t,t+dt]$,
where the deterministic function 
$F:\Omega\rightarrow (0,\infty)$
is an injection which defines the image of 
the process
$F_t$.

Our aim in this section is to extend (i.e. lift) the generator
$\L$
to the generator 
$\widetilde{\L}$,
which will describe the dynamics of the lifting 
$(F_t, I_t)$
of our original process
$F_t$.
The component
$I_t$
will be a finite-state continuous-time Markov chain, which will be 
described shortly, that approximates the
realized variance, 
up to time 
$t$,
of the underlying process
$F_t$.
The generator 
of the chain
$(F_t, I_t)$
will give us the probability kernel
for the process 
$I_t$,
which is what we are ultimately interested in
as it contains the probability distribution 
of the relevant path information.

We will start by describing a continuous process
$\Sigma_t$
which will contain the relevant path-information of the underlying 
Markov chain
$F_t$
and then define a finite-state stochastic process
$I_t$
that will be used to approximate 
$\Sigma_t$. 
The key feature of the state-space of the process 
$I_t$,
which allows numerical tractability of our approach,
is that it is contained on a circle in the complex plane
(see Section~\ref{sec:pricing}).

The procedure we are about to describe works for a specific type of 
path-dependence only. 
Assume that if at time 
$t$
the underlying process 
$F_t$
is at a state 
$F(x)$
for some
$x\in\Omega$,
then the change of the value 
$\Sigma_t$,
in the infinitesimal time interval
$dt$
is of the form
$$d\Sigma_t=Q(x)dt,$$
where the function
$Q:\Omega\rightarrow\RR$,
defined in terms of the underlying stochastic process,
has two key properties:
\begin{itemize}
\item the mapping 
      $Q$
      is independent of the path the 
      underlying process follows on the interval
      $[0,t)$
      and
\item the value 
      $Q(x)$
      depends on the level 
      $x$
      at time
      $t$
      and on the distribution of the underlying 
      process in the infinitesimal future
      time interval
      $dt$
      as given by the Markov generator
      $\L$.
\end{itemize}

The first property states that the change in the path-information
$\Sigma_t$,
over a short time period 
$[t,t+dt]$,
does not depend on the path
taken by the underlying process
up to time 
$t$.
The second property tells us that the evolution of the path information
over the interval
$[t,t+dt]$,
conditional on the current level of the process
$F_t$,
is determined by the future distribution
of
$F_t$
over the infinitesimal time interval.

It is clear that the realized variance of the process
$F_t$
up to time
$t$,
can be captured by a random process 
$\Sigma_t$.
Indeed, if we define the function
$Q$
in the following way
\begin{eqnarray}
\label{eq:InstVar}
Q(x):=\sum_{y\in\Omega}\left(\frac{F(y)-F(x)}{F(x)}\right)^2\L(x,y)
\end{eqnarray}
it follows immediately that the above conditions are satisfied. 
This is because the realized variance of
$F_t$
is simply an integral over time of the instantaneous variance of
$F_t$
which is given by~(\ref{eq:InstVar}).

The last observation is crucial for all that follows, because it implies
that the process 
$\Sigma_t$
is uniquely determined by its state-dependent 
instantaneous drift.
In particular we see that the process
$\Sigma_t$
has no volatility term and that it has continuous 
sample paths since it allows a representation
as an integral over time.

The fundamental consequence of these facts is 
the following:
a finite-state Markov chain 
$I_t$
can be used as a model
for the process
$\Sigma_t$
if and only if 
the instantaneous drift of the chain 
is equal to
$Q(x)$,
whenever the underlying process
$F_t$
is in state
$F(x)$,
and the instantaneous variance of 
$I_t$
is equal
to zero up to first order.
The first requirement clearly follows from the discussion above.
The second condition is there to reflect the fact that 
the process
$\Sigma_t$
is a continuous It\^{o} process which
has no volatility term. A non-constant random process 
on a lattice will always have a non-zero instantaneous
variance, but the second condition ensures that this 
instantaneous variance 
goes to zero as quickly as  
the lattice spacing itself.
In other words the Markov chain approximation 
$I_t$
for the process
$\Sigma_t$
must exhibit 
neither diffusion nor jump behaviour.
Therefore,
as we shall see in the next subsection,
$I_t$ 
must be a Poisson process 
with state-dependent intensity. 
 

\subsection{Lifting of the generator for the underlying process} 
\label{subsec:Lifting}
Let
$I_t$
as above 
denote the Markov chain whose value 
approximates the realized variance 
$\Sigma_t$
of the 
forward price
$F_t$
from time 
$0$
to time
$t$.
We shall express
$I_t$
as
$\alpha m_t$
where 
$m_t$
is a Poisson process (with non-constant intensity)
starting at 
$0$
and gradually jumping up along the grid 
given by
$\Psi=\{0,\ldots,2C\}$,
where 
$C$
is an element in
$\NN$.
The constant 
$\alpha$
controls the spacing of the grid for the realized variance
$I_t$.

We are now going to specify precisely the dynamics
of the process
$m_t$
which is
a fundamental ingredient of our model. As mentioned before, the process
$m_t$
will behave as a Poisson process whose intensity is determined
by the level of the underlying. In other words the 
Markov generator, conditional on the underlying
process being at the level
$F(x)$, is of the form

\begin{eqnarray}
\label{eq:VarGen}
 \L^{m}(x:c,d) := \left\{\begin{array}{ll}
                       \frac{1}{\alpha}Q(x) & \mathrm{if}\>\>d=(c+1)
                                                \!\!\!\mod (2C+1);\\
                                             & \\
                       -\frac{1}{\alpha}Q(x) & \mathrm{if}\>\>d=c. 
                          \end{array}
                        \right. 
\end{eqnarray}
The variables
$c$
and
$d$
are elements of the discrete set
$\Psi=\{0,\ldots,2C\}$
and the function 
$Q$
is the instantaneous variance of the underlying 
process
as defined in~(\ref{eq:InstVar}).
%
This family of generators specifies the dynamics of 
the process
$I_t$
with the following instantaneous drift
\begin{eqnarray}
\lim_{\Delta t\rightarrow 0}
\EE\left[\frac{I_{t+\Delta t}-I_t}{\Delta t}\Bigl\lvert F_t=F(x),I_t=I(c)
 \right]& = & \sum_{d=0}^{2C}(I(d)-I(c))\L^{m}(x:c,d) \nonumber\\
\label{eq:InstDrift}        
        & = & \alpha(c+1-c)\frac{1}{\alpha}Q(x)=Q(x),
\end{eqnarray}
for all values
$F(x)$
of the underlying process
$F_t$
and all integers
$c$
which are strictly smaller than 
$2C$.
This implies that the Markov chain
$I_t$
has the same instantaneous drift
as the actual realized variance.
A similar calculation tells us that the 
instantaneous variance of
$I_t$
equals
$\alpha Q(x)$.
Since 
$\alpha$
is the spacing of the lattice for the chain
$I_t$,
we have just shown that the first two instantaneous 
moments of
$I_t$
match the first two moments of the realized variance
$\Sigma_t$
for all points on the lattice
$\{0,\ldots,2C\}$
except the last one. 

Notice however that the equality~(\ref{eq:InstDrift}) 
breaks down if
$c$
equals
$2C$,
because we have imposed periodic boundary condition
for the process
$I_t$.
Put differently this means that 
$I_t$
is in fact a process on a circle
rather than on an interval. The latter
would be achieved if we had imposed absorbing 
boundary conditions
at
$2C$. 
That would perhaps be a more
natural thing to do
since the process 
$\Sigma_t$
is certainly not periodic.
But an absorbing 
boundary condition would destroy the 
delicate structure of the spectrum
of the lifted generator
$\widetilde{\L}$
which is preserved by the periodic 
boundary condition.
It is precisely this structure 
that makes 
the periodic nature
of
$I_t$
a key ingredient of our model,
because it allows us to linearize 
the complexity of the pricing algorithm
(see appendix~\ref{sec:Bloc}).
It should be noted that the general philosophy 
behind either choice of the boundary condition
would be the same: the lattice in the calibrated model 
should be set up in such a way that the process
never reaches the boundary value
$2C$,
because if it does an inevitable loss of information
will ensue regardless of the boundary conditions we choose.

We are now finally in a position
to define the lifted Markov generator of
the process
$(F_t,I_t)$
as
\begin{eqnarray}
\label{eq:lift}
\widetilde{\L}(x,c;y,d) := \L(x,y)\delta_{c,d}
                                 + \L^m(x:c,d)\delta_{x,y}, 
\end{eqnarray}
where
$x,y$
are in the state-space of the chain 
$F_t$
and
$c,d$
are elements in
$\Psi$.
The structure of the spectrum of the operator
$\widetilde{\L}$
will be exploited in Section~\ref{sec:pricing}
to obtain a pricing algorithm for payoffs
which are general 
functions of the realized variance
(see Theorem~\ref{thm:LiftKer}).
The reason for
specifying the generator
$\L^m(x:c,d)$
by~(\ref{eq:VarGen}) (i.e. insisting on 
the periodic boundary condition for the process
$I_t$)
will become clear in 
the proof of Theorem~\ref{thm:LiftKer},
which exploits 
the spectral properties 
of the lifted generator
$\widetilde{\L}$
(see appendix~\ref{sec:Bloc}
for the precise description of the
spectrum of
$\widetilde{\L}$).

\section{Pricing of derivatives on realized variance}
\label{sec:pricing}
Recall that the generator 
$\L$
for the underlying continuous-time Markov
chain
$F_t$ 
in the forward measure,
described in~\cite{FXrev},
is defined on the state-space
$\Omega\times V$
via an injective function
$F:\Omega\times V\rightarrow (0,\infty)$.
The set 
$\Omega$
with 
$N$
elements is the state-space
of the various jump-diffusions and the set
$V$
with 
$M$
elements is the state-space for the random switching 
between regimes
(see the formulae on page 7 in~\cite{FXrev} for the precise 
definition of the generator matrix
$(\L(x,\beta;y,\gamma))_{(x,\beta),(y,\gamma)\in \Omega\times V}$).
In Section~\ref{sec:calibration} we shall see that
for the options data on the S\&P 500 used in this paper, 
it suffices to take 
$N=76$
with only two regimes
($M=2$).


Let
$\Sigma_T$
be the realized variance over the time interval
$[0,T]$,
expressed in annual terms,
as defined in~(\ref{eq:DiscreteVar}).
In this section we are going to find pricing formulae
for general payoffs that depend on the annualized realized variance
$\Sigma_T$.

Our first task is to find the probability kernel
of the lifted process
$(F_t,I_t)$
which was defined in Section~\ref{sec:path}.
Recall that the Markov chain 
$I_t=\alpha m_t$,
which is used to model 
the realized variance,
is specified in terms of a translation
invariant process
$m_t$,
whose domain is 
$\Psi=\{0,\ldots,2C\}$,
and some positive constant 
$\alpha$
which determines the lattice spacing for the domain of
$I_t$.
The value
$C\in\NN$
specifies the size of the lattice for the realized
variance and has to be large enough so that
the process 
$I_t$,
starting from zero,
does not transverse the entire lattice.
This is a very important technical requirement
as it ensures that there is no
probability leakage in the model
(which is theoretically possible since we are using 
periodic boundary conditions
for the process
$I_t$).
The dynamics of the chain
$m_t$
are given by the Markov generator in~(\ref{eq:VarGen}).

Recall also that the process
$I_t$
records the total realized variance up to time 
$t$,
which implies that 
the annualized realized variance
$\Sigma_T$
that interests us,
will be described by  
$\frac{1}{T}I_T$
where the time horizon
$T$
is expressed in years. The key ingredient in the calculation 
of the probability distribution function of the process
$(F_t, I_t)$
is the block-diagonalization algorithm from 
subsection~\ref{subsec:alg} of the appendix. We are now going to apply it
to the generator~(\ref{eq:lift})
in order to find the joint pdf of the lifted process.


\subsection{Probability kernel of the lift ($\mathbf{F_t, I_t}$)}
The generator
$\widetilde{\L}(x,\beta,c;y,\gamma,d)$
of the process
$(F_t, I_t)$,
given by~(\ref{eq:lift}),
is a square partial-circulant matrix
as defined in appendix~\ref{sec:Bloc}.
It is also clear that the dimension  of the vector space
acted on by the matrix 
$\widetilde{\L}$
is
$MN(2C+1)$.
The coordinates 
$(x,\beta,c),(y,\gamma,d)$
of the matrix
$\widetilde{\L}$
(i.e. the lattice points of the process)
lie in the set
$\Omega\times V\times\Psi$,
where
$\Omega$
is the grid for the underlying forward rate and the set
$V$
contains all volatility regimes of the model
for the forward
$F_t$.
Notice that the circulant matrices
$\L^m$
from~(\ref{eq:VarGen}),
used in the definition of 
$\widetilde{\L}$,
are very simple because the only non-zero elements are
on the diagonal, just above the diagonal and the element 
in the bottom left 
corner of the matrix. 
In other words
if we interpret the matrix
$\L^m$,
associated with the lattice point
$(x,\beta)$,
in terms of the definition of a circulant matrix
given at the beginning of appendix~\ref{sec:Bloc},
we see that 
$$c_1=-c_0=\frac{Q(x,\beta)}{\alpha},$$
where the function
$Q(x,\beta)$
is the instantaneous variance as defined by~(\ref{eq:InstVar})
and
the constant
$\alpha$
is the lattice spacing of the domain of
$I_t$.
All other elements 
$c_j$
are equal to zero.

It is therefore clear that,
using the expression (\ref{eq:CirculantEigValue})
for the eigenvalues of circulant
matrices,
equation~(\ref{eq:BlocDiagForm})
can be reinterpreted as
$$
\L_k(x,\beta;y,\gamma) 
:= \L(x,\beta;y,\gamma) + \delta_{(x,\beta),(y,\gamma)}\left(e^{-ip_k}-1\right)\frac{Q(x,\beta)}{\alpha},
$$
where
$\L_k$
is the 
$k$-th 
block in the block-diagonal decomposition of
$\widetilde{\L}$
and
the value of 
$p_k$
is given by the expression
\begin{eqnarray}
\label{eq:pk}
p_k := \frac{2\pi}{2C+1}k.
\end{eqnarray}
The index 
$k$
in these expressions runs from 
$0$
to
$2C$.
Notice that the matrices
$\L_k$
differ from the Markov generator
$\L$
only along the diagonal.  
We are now in a position to state the key
theorem that will allow us to find the probability
kernel of the lifted process.

\begin{thm}
\label{thm:LiftKer}
Let
$\widetilde{\L}$
be the Markov generator of the stochastic process
$(F_t,I_t)$
as described in Section~\ref{sec:path}
and let
$\phi:\CC\rightarrow\CC$
be a holomorphic function.
Then the following equality holds 
$$
\phi(\widetilde{\L})(x,\beta,c;y,\gamma,d) = \frac{1}{2C + 1} 
\sum_{k=0}^{2C} e^{-ip_k (c-d)} \phi(\L_k)(x,\beta;y,\gamma),
$$
where 
$\L_k$
is the operator defined above,
$p_k$
is given by~(\ref{eq:pk})
and 
$(x,\beta,c),(y,\gamma,d)$
are elements
of
$\Omega\times V\times\Psi$.
\end{thm}

Before embarking on the proof of this theorem we may
summarize as follows: if a
linear operator 
$A$
can be block-diagonalized 
by a 
discrete Fourier transform
(cf. last paragraph of subsection~\ref{subsec:alg}
in the appendix), 
then so can any operator
$\phi(A)$
where 
$\phi$
is a holomorphic function defined on the entire
complex plane. Note that the assumptions on 
function 
$\phi$
and the linear operator
$\widetilde{\L}$
in Theorem~\ref{thm:LiftKer}
are too stringent and in fact the theorem holds 
in much greater generality. 
For our purposes 
however 
the setting 
described in the theorem is sufficient 
as it applies directly to the model.
Therefore the proof of the theorem, given in appendix~\ref{subsec:proof},
only applies in
the restricted case stated above.
Note that the idea of semi-analytic diagonalization, which 
is central to the proof of our theorem, has been applied extensively in
physics and numerical analysis (see for example~\cite{Haldane},
\cite{Bavely},
\cite{Dongarra}
and the references therein).

\begin{figure}[ht]
\includegraphics[width=11cm]{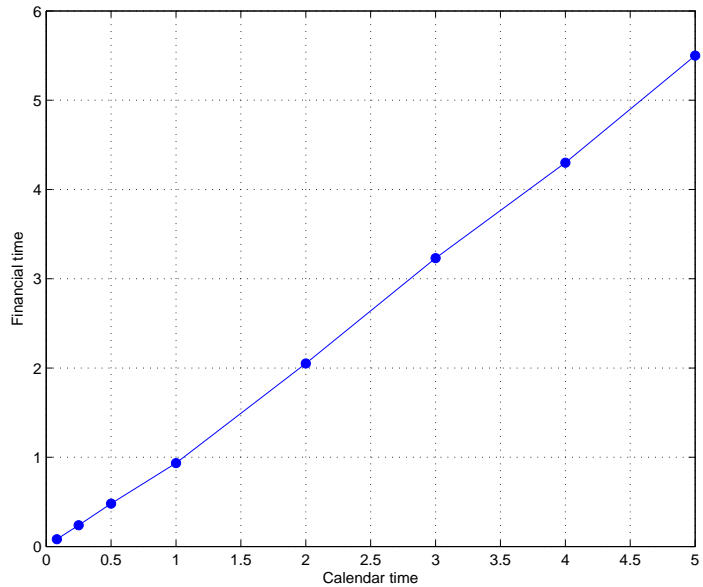}
\caption{\footnotesize{Deterministic time-change $f(t)$ 
                       (measured in years) 
                       as a function of calendar
                       time $t$ (also in years). Function
                       $f$
                       was used in the calibration of the model 
                       to the vanilla surface 
                       of the S\&P 500.}}
\label{fig:timechange}
\end{figure}

We can now find the full probability kernel
of the process 
$(F_t,I_t)$
by applying Theorem~\ref{thm:LiftKer}
in the following way.
More explicitly, for any pair of calendar times
$t$
and
$T$,
such that 
$t<T$, 
let
the conditional probability
$\PP(F_T=F(y,\gamma),I_T=d \lvert F_t=F(x,\beta),I_t=c)$
be denoted by
$p((x,\beta,c),t;(y,\gamma,d),T)$.
The well-know exponential solution to the 
backward Kolmogorov equation and Theorem~\ref{thm:LiftKer}
imply 
\begin{eqnarray}
p((x,\beta,c),t;(y,\gamma,d),T) & = &
e^{(f(T)-f(t))\widetilde{\L}}(x,\beta,c;y,\gamma,d) \nonumber \\
                               & = & \frac{1}{2C + 1} 
\sum_{k=0}^{2C} e^{-ip_k (c-d)} e^{(f(T)-f(t))\L_k}(x,\beta;y,\gamma) \nonumber\\
\label{eq:FinalPdf}
                               & = & \frac{1}{2C + 1}
                               \sum_{k=0}^{2C} 
\sum_{n=1}^{M\cdot N} e^{\lambda^k_n (f(T)-f(t))} e^{-ip_k (c-d)} 
u^k_n(x,\beta) v^k_n(y,\gamma),
\end{eqnarray}
where 
$\lambda^k_n$
are the eigenvalues 
and
$u^k_n$
are the eigenvectors
of 
$\L_k$.
As usual we denote by
$v^k_n$
the columns of the matrix
$U_k^{-1}$,
where 
$U_k$
consists of all the eigenvectors 
of
$\L_k$.
Function
$f$
in the above formula is the deterministic time-change 
for the underlying model which was
introduced in~\cite{FXrev}, subsection~2.4 (see 
Figure~\ref{fig:timechange} for the specification
of function
$f$
implied by the volatility surface for the S\&P 500
used in Section~\ref{sec:calibration}).

Formula~(\ref{eq:FinalPdf}) is our key result because it allows 
us to price any derivative which depends jointly
on the realized variance of the underlying index and the index
itself at any time horizon
$T$.
In the following subsection
we state an explicit formula for the value of such a derivative.

\subsection{Pricing derivatives on the realized variance}
\label{subsec:PricingVar}

Let us assume that we are given 
a general payoff 
$h(\Sigma_{T-t})$
that depends on the annualized realized variance
$\Sigma_{T-t}$
between
the current
time
$t$
and some future expiry date
$T$.
The current price of a derivative with this payoff
within our model can be computed directly using
the joint probability distribution function~(\ref{eq:FinalPdf})
in the following way
\begin{eqnarray*}
\label{eq:VarDerPrice}
C_t(x,\beta)=e^{-(r(T)T-r(t)t)}
\sum_{d\in\Psi}
\left(\sum_{(y,\gamma)\in\Omega\times V}
p((x,\beta,0),t;(y,\gamma,d),T)\right)h\left(\frac{\alpha d}{T-t}\right).
\end{eqnarray*}

The sum in the brackets is the marginal distribution
of the process
$I_t$
at time 
$T-t$.
Notice that the realized variance process must always start from
$0$
at the inception of the contract.
The factor
$\frac{1}{T-t}$
normalizes the value of
$I_{T-t}$
so that it is expressed in annual terms.
As before, the constant 
$\alpha$
is the lattice spacing for the realized variance.

The point 
$x$
from
$\Omega$
is chosen in such a way that the equation
$F(x)= e^{(r(t)-d(t))t}S_t$
holds,
where 
$S_t$
is the index level at the current time  
$t$,
the functions 
$r,d:[0,T]\rightarrow\RR_+$
are the deterministic instantaneous interest rate and
dividend yield respectively 
and 
$\beta$
in
$V$
corresponds to the volatility regime at time
$t$.

Notice that because in the valuation formula above there is no
restriction on the payoff function
$h$,
the contract 
$C_t$
can be anything from a variance swaption to a volatility swap
(or swaption) 
and can be priced very efficiently using the calibrated model.
Since formula~\eqref{eq:FinalPdf}
yields the joint distribution of the realized variance 
$\Sigma_{T-t}$
and
the underlying security
$S_T$,
a trivial modification of the above  expression 
would give us the current price 
in our model of any payoff
of the form
$h(S_T,\Sigma_{T-t})$.

\section{Calibration to the vanilla surface}
\label{sec:calibration}
\begin{figure}[ht]
        \includegraphics[width=11cm]{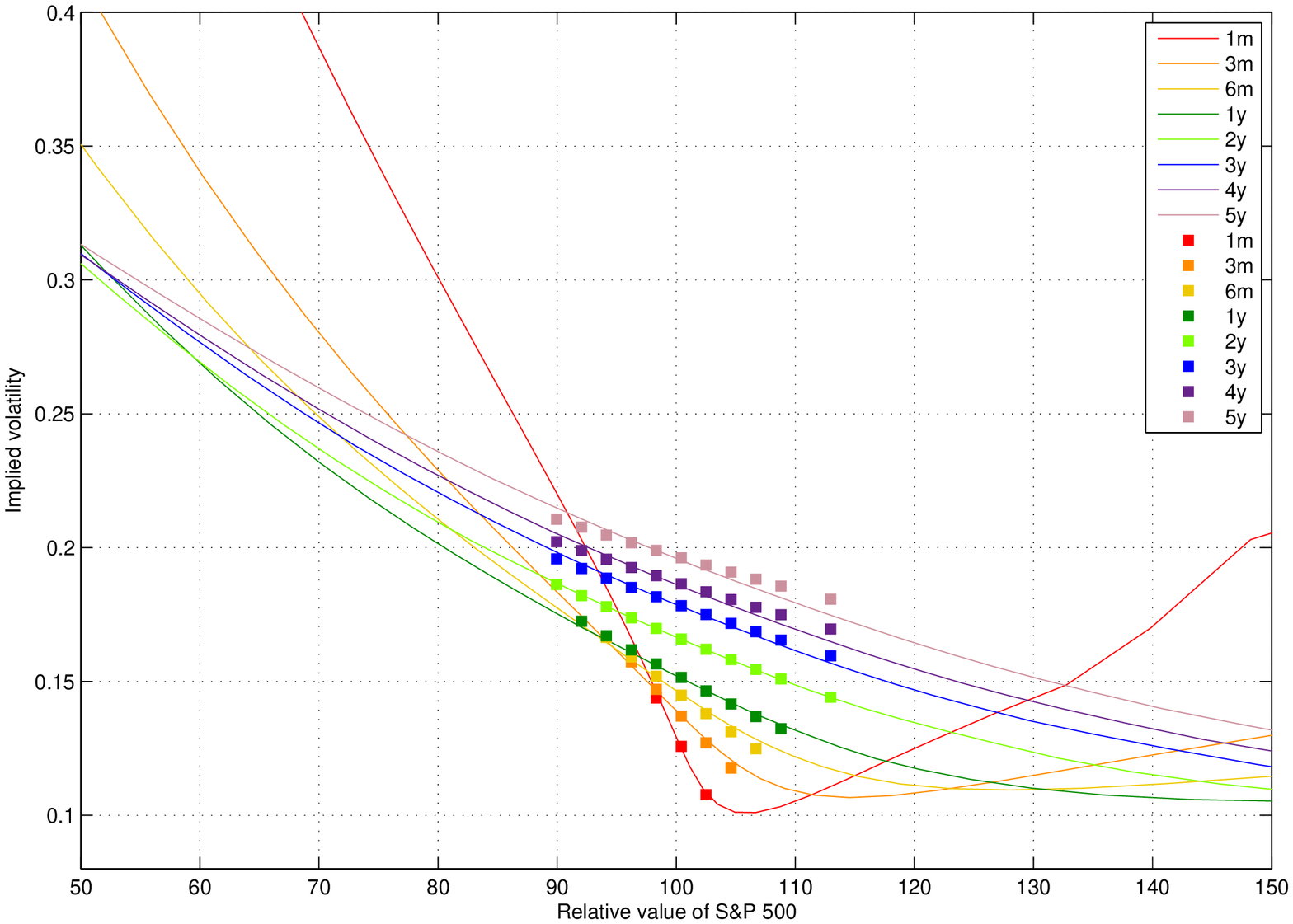}
        \caption{\footnotesize{Implied volatilities 
                               for out-of-the-money European options
                               written on the 
                               S\&P 500 with
                               maturities between 3 months and 
                               10 years. 
                               For each market-specified maturity the 
                               squares
                               $\blacksquare$
                               represent market-implied volatilities. 
                               The continuous
                               curves graph the implied volatility of 
                               the model as a function of strike.
                               The relative value of S\&P 500, with respect
                               to the current level of spot,
                               is plotted along the line
                               of abscisse.}}
    \label{fig:ivols}
\end{figure}

We are now going to 
calibrate the model for the underlying, described in Section 2
of~\cite{FXrev},
to the implied volatility surface of the S\&P 500 equity index
for maturities between 1 month and 5 years
(see legend of Figure~\ref{fig:ivols}
for all market defined maturities used in the calibration) and 
a broad range of liquid strikes for each maturity.
The market data consist of the implied
Black-Scholes volatilities
for each strike and 
maturity.\footnote{Notice that we are using more strikes
for longer maturities than for shorter maturities.
This is not an inherent requirement but a consequence
of the initial structure of the market data we used for calibration 
of the model.}
Our task is to find a set of values for the parameters 
of the model such that if we reprice the above options 
and express their values in terms of the implied
Black-Scholes volatilities,
we reobtain the market quotes.

The first choice we need to make pertains to the number of regimes
$M$
we are going to use. 
In foreign exchange markets the implied volatility 
skew exhibits complex patterns
of behaviour across currency pairs and
the corresponding model required no fewer than 4 regimes to accommodate this
complexity
(see~\cite{FXrev}). 
In equity markets, by contrast, the skew is always
large and negative and becomes less pronounced as time to
maturity increases (see chapter 8 in~\cite{Gatheral}).  
Thus in the case of S\&P 500 two regimes prove sufficient 
($M=2$).
The number of regimes cannot be smaller than two because
in the case of a single regime the model becomes a 
jump diffusion which, as is well-known, does not describe
adequately the implied volatility surface (see chapter 5
in~\cite{Gatheral}).

In order to calibrate our model 
we select an non-homogeneous grid with
$76 = N$ 
points
used to span the possible values of the forward
rate
$F_t^\Omega$.
The grid is elliptical in the following sense: close to the
current value of the spot, the lattice points
are very dense; the density goes down as we go further away 
from the current level of spot. The top node is at
$100^2$ 
and the bottom node at
$100^{-2}$.

The underlying grid for the model is therefore of the size
$MN= 2\cdot76=152 $,
which makes the model more efficient computationally than the
one used for modelling the foreign exchange
rate, where the lattice was roughly 
double in size.

The next group of model parameters which do not need to be changed
with every calibration are the levels 
$F_0$
and
$F_1$
and the stochastic volatility matrices
$\G_{0}^V$
and
$\G_{1}^V$.
Since the model is defined so that the starting regime is
regime 1, we take the level 
$F_1$
to be equal to the current value of spot 100.
The level  
$F_0$
is chosen somewhat arbitrarily to be 95. If the underlying
starts trading around this level, regime 0 assumes a more
dominant role in the model. If we took
$F_0$ 
to be, for example, 90,
that would have a minor effect on the short term (up to six months)
implied volatility skew of the model and a negligible
effect on longer maturity skews.
This is because by moving
$F_0$
from 95 to 90 we decrease the influence of regime 1
on the short term maturity skew.

The stochastic volatility generators 
$\G_{0}^V$
and
$\G_{1}^V$
are chosen to reflect the fact that in equity markets, a downward
move in the underlying index results in steeper negative skews. 
The starting regime 1 corresponds to the current skews
for short maturities,
whereas regime 0 corresponds to steeper negative skews, which will occur
if the value of the underlying index drops.
The generator
$\G_{0}^V$,
which is dominant if the index is trading around 95 or below,
assigns a very slight probability to changing back to regime 1.
This is because if the index is trading at 95 or below, 
the skew is unlikely to become less steep.
For the same reason 
$\G_{0}^V$
assigns a substantial probability to staying in regime 0. 
On the contrary, the generator
$\G_{1}^V$
assigns roughly the same probability to staying in regime 1
and to switching to regime 0. This is because when the underlying
index is trading close to the current level of spot, the skews
can become less or more steep with roughly the same probability.

In order to calibrate to the specific instance of the volatility 
surface, the parameters in the CEV jump diffusion need to be chosen.
Note that we can calibrate the model 
to the entire implied volatility surface
without recourse to the
parameters
$\bar\sigma_\alpha$,
for 
$\alpha=0,1$.
This is due to the fact that in our model
they only affect option values 
for strikes below 20. 
Since, when calibrating to the implied volatility
surface, we are not interested in strikes that are so far away from
the at-the-money strike, we could take the value of the parameters
$\bar\sigma_\alpha$,
for 
$\alpha=0,1$,
to be 
$10000\%$
(this choice amounts to excluding the paramters
$\bar\sigma_\alpha$,
for 
$\alpha=0,1$,
from the model).

On the other hand, the values of variance swaps are very sensitive
to the strikes below 20
(see expression~(\ref{eq:VarHedge})
in appendix~\ref{subsec:VIX}
for the hedging portfolio consisting of vanilla options)
and therefore exhibit a strong dependence on the
$\bar\sigma_\alpha$.
Therefore if we set
$\bar\sigma_\alpha=10000\%$,
for 
$\alpha=0,1$,
variance swaps would be grossly overpriced.
Since parameters 
$\bar\sigma_\alpha$,
for 
$\alpha=0,1$,
do not affect the at-the-money skew,
we could use these parameters to calibrate the model 
to the term structure of variance swaps.
In this paper we did not have the relevant variance swap
data to calibrate to, we therefore chose 
$\bar\sigma_0=\bar\sigma_1=60\%$
which gave reasonable values for the term structure 
of variance swaps
(see Figure~\ref{fig:VarSwapTerm}). 
Note also that
$\bar\sigma_1$
influences more markedly the short term variance swaps and
$\bar\sigma_0$
has a greater effect on longer term contracts.
Therefore the steepness of the term structure
of variance swaps implied by the model 
can be adjusted by taking different values for
$\bar\sigma_1$ 
and
$\bar\sigma_0$.

Going back to the remaining parameters of the CEV jump diffusion,
note that, since there are no positive jumps in the equity index,
we take
$\nu^+_\alpha$,
for 
$\alpha=0,1$,
to be 0.
We take the CEV volatility 
$\sigma_1$
in regime 1
to be close to the at-the-money implied volatility
for the shortest maturity we are calibrating to (in this case
one month). The parameters 
$\beta_1$
and 
$\nu^-_1$
are chosen to fit the skew for the maturity of one month.

Introducing a second regime into the model
does not alter the one-month skew
(this follows from the definition of the 
stochastic volatility generator
$\G_{1}^V$).
Since regime 0 corresponds to the underlying index trading
at a lower level, the values of
$\beta_0$
and
$\nu^-_0$
are more extreme, as they correspond to the tightening of the
skew. Their values are chosen in such a way that the model qualitatively
fits the skews for all maturities (i.e. the risk-reversals are priced
correctly but the level of the at-the-money implied volatility
is not necessarily matched).

By choosing the parameters using these guidelines
(see table~\ref{t:para}), 
we obtain an implied volatility surface that has the correct 
qualitative shape, but the at-the-money levels are not represented
correctly. To remedy this, mild explicit time-dependence needs
to be introduced. This is achieved by introducing a deterministic   
function which transforms calendar time to financial time (see
Figure~\ref{fig:timechange}). The effect on the implied
volatility skew is that of a parallel shift for each maturity.
In other words, the shape of the skew remains unchanged.  
For the final result see Figure~\ref{fig:ivols}.

The main calibration criterion has been to minimize the
explicit time-dependence in order to preserve the 
correct (i.e. market implied) skew through time.
The stationarity requirement in the calibration ensures that
the forward skew\footnote{There is a closed
form solution for the value of a forward-start in the Black-Scholes
model and the only unknown parameter in that formula 
is the forward volatility.
It is therefore 
customary to define a forward smile of any model as a function
mapping the forward strike to the implied forward
volatility which is obtained by inverting the Black-Sholes formula
(see
subsection~\ref{subsec:VIX} for the precise definition
of a forward strike).}
for any maturity retains the desired shape as can 
be observed in Figures~\ref{fig:fs_skew3m}, \ref{fig:fs_skew6m},
\ref{fig:fs_skew1y} and~\ref{fig:fs_skew2y}.


The S\&P 500 index equaled 1195 at the time when the option data were
recorded. Throughout the paper a relative value
of the index, set at 100, is used for simplicity.
The forward
price levels 
$F(x)$,
where
$x$
is an element
of the underlying 
grid
$\Omega$,
are also measured on the relative scale.
As mentioned above our starting volatility regime is
regime 
$\alpha=1$.

\begin{table}[ht]
\begin{center}
\begin{tabular}{|c|c|c|c|c|c|c|} \hline $\alpha$ & $\sigma_\alpha$ &
$\beta_\alpha$ & $\bar\sigma_\alpha$ & $\nu^-_\alpha$ &
$\nu^+_\alpha$ & $F_\alpha$\\
\hline
0 & 16\%  & -0.8  & 60\%  & 0.18 & 0 & 95.00 \\
1 & 13\%  & -0.3  & 60\%  & 0.15 & 0 & 100.00\\
\hline
\end{tabular}
\end{center}
\vspace{2mm}
\caption{Parameters for the local volatility regimes and the jump intensities.}
\label{t:para}
\end{table}

\begin{table}[ht]
\[
\G_{0}^V=
\begin{pmatrix}
-1 & 1 \\
5 & -5
\end{pmatrix}, \>\>\> 
\G_{1}^V=
\begin{pmatrix}
-5 & 5\\
 7 & -7
\end{pmatrix} 
\]
Markov generators for stochastic volatility ($\alpha=0, 1$).
\end{table}

The code used to obtain these results 
is written in VB.NET
and relies on LAPACK
for the computation of the spectrum
of the generator.
In the current model the generator
is a matrix of the size
$152\times152$.
The time required to diagonalize
the generator and price
76 options per maturity
for 6 maturities (i.e. 456 
options) is less than 4 seconds
on a single Pentium M processor
with 1GB of RAM.
Note that the time consuming step 
is the computation of the spectrum
and eigenvectors. Once these are known
the probability kernel for any maturity,
and therefore all option prices for that maturity,
can be obtained within a fraction of a 
second.

\subsection{The parameters $C$ and $\alpha$ in the lift ($F_t,I_t$)}
\label{subsec:CandAlpha}
\begin{figure}[ht]
\centering\noindent
       \includegraphics[width =0.47\linewidth]{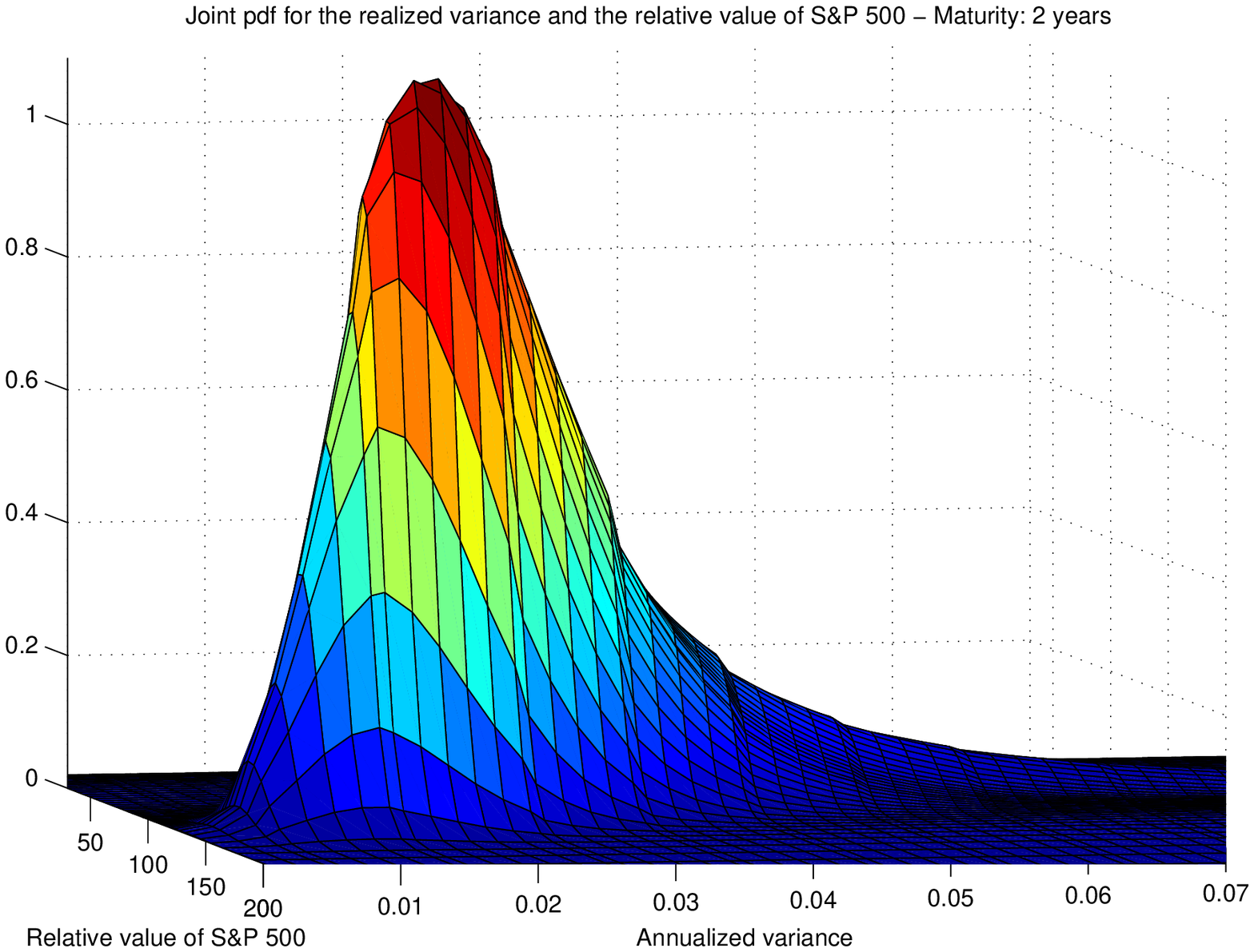}
\hfill
\includegraphics[width =0.47\linewidth]{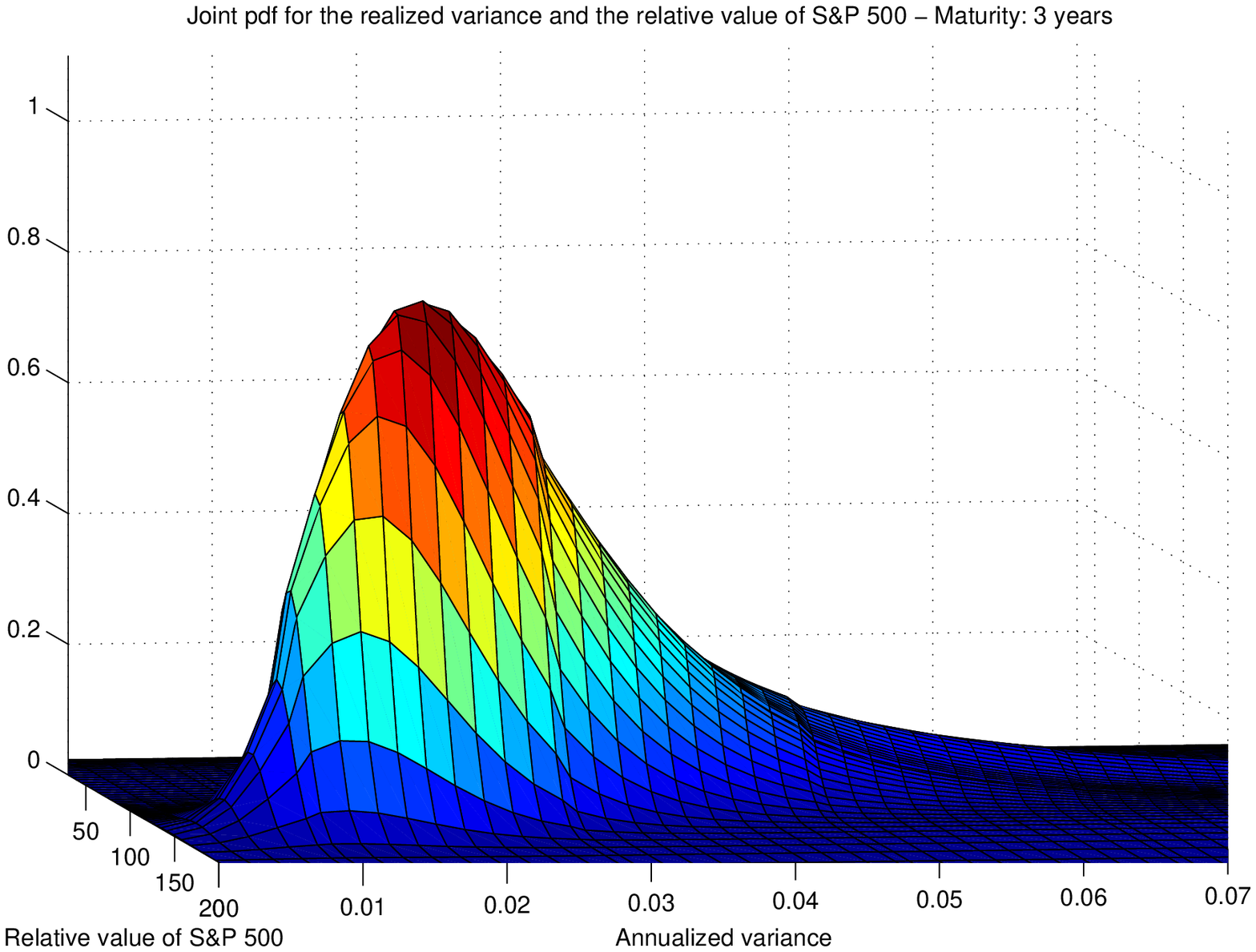}
\caption{\footnotesize{Joint probability distribution function 
                       for the annualized realized volatility 
                       and the spot rate of S\&P 500 in 2 years' and 3 
                       years' time.}}
\label{fig:Joint_2y}
\end{figure}

\begin{figure}[ht]
\centering\noindent
       \includegraphics[width =0.47\linewidth]{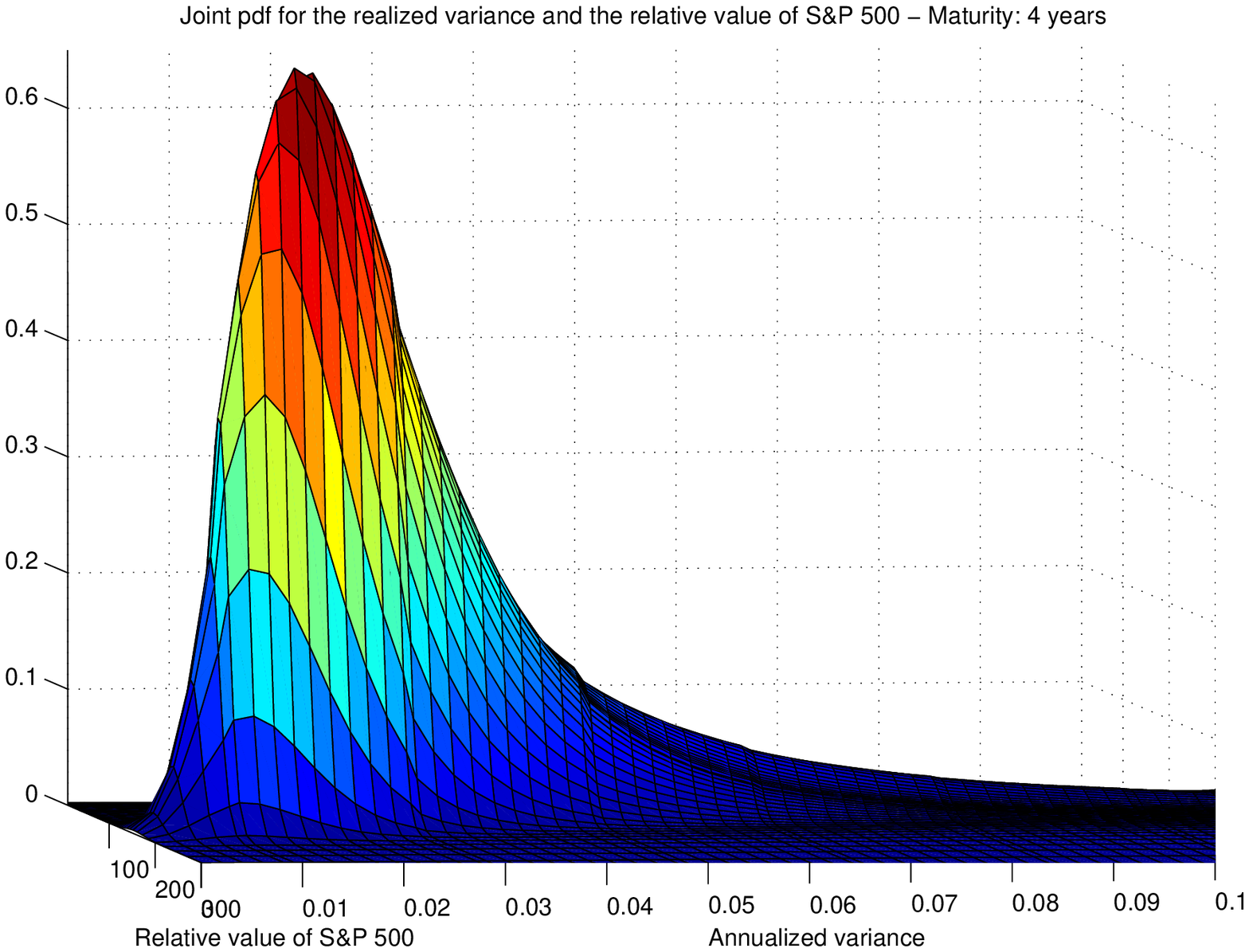}
\hfill
\includegraphics[width =0.47\linewidth]{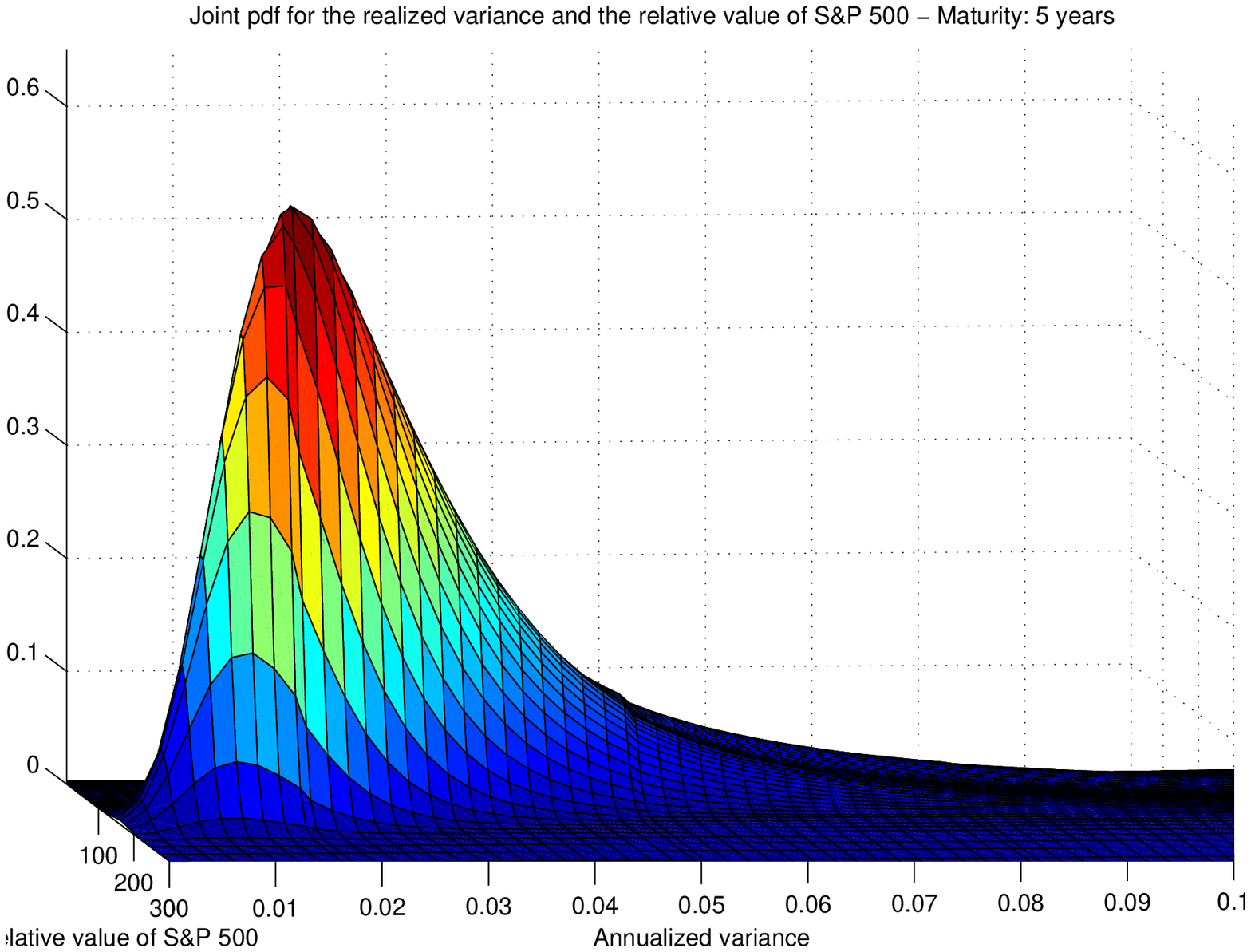}
\caption{\footnotesize{Joint probability distribution function 
                       for the annualized realized volatility 
                       and the spot rate of S\&P 500 in 4 years' and 
                       5 years' time.}}
\label{fig:Joint_4y}
\end{figure}

Figures~\ref{fig:Joint_6m}, 
\ref{fig:Joint_2y}, 
and~\ref{fig:Joint_4y}
contain the graphs of the joint distribution functions of
the spot level and the annualized realized variance 
for maturities between 
6 months and 
5 years.
The marginal distributions 
of the annualized realized variance, obtained from the 
above joint distributions by integration
in the dimension of 
the spot value
of the index,
are shown in  
Figure~\ref{fig:RealizedVar}.

\begin{figure}[ht]
\centering\noindent
       \includegraphics[width=11cm]{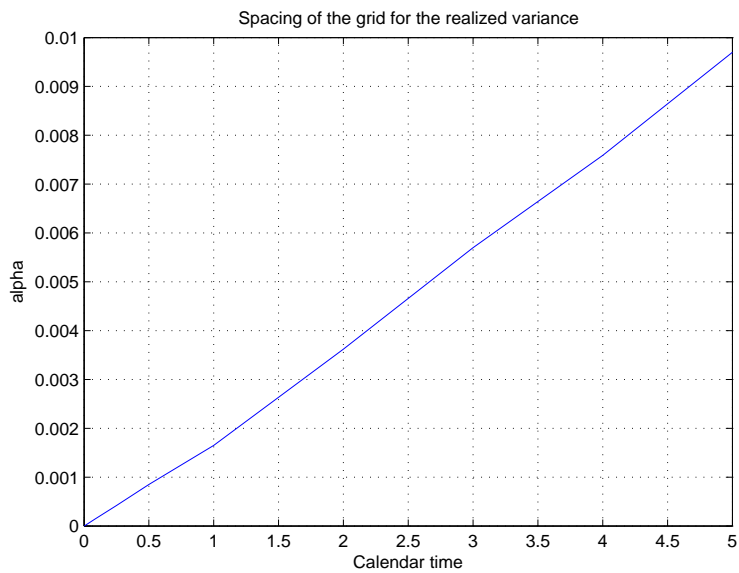}
\caption{\footnotesize{The spacing of the grid 
                       $\alpha(t)=\frac{0.42^2}{C}f(t)$  
                       for the realized 
                       variance as a function of calendar time
                       $t$.
                       The constant 
                       $C$
                       equals 
                       100,
                       for all times 
                       $t$
                       from now to 5 years,
                       and corresponds to 201 states for the 
                       process 
                       $I_t$
                       that represents the realized variance
                       of the underlying  
                       up to time 
                       $t$.
                       The function
                       $f(t)$
                       represents financial time and is graphed
                       in Figure~\ref{fig:timechange}.}}
\label{fig:Alpha}
\end{figure}

The parameters in the model that influence the 
dynamics of the realized variance 
$I_t$
are
the number of lattice points 
$C$
and the lattice spacing 
$\alpha$.
It turns out that for maturities
up to 5 years it is sufficient
to take
$201=2C+1$
uniformly distributed points for the grid of the realized
variance
(i.e.
$C=100$).
Table~\ref{t:VarSpacing}
contains the values 
of
$\alpha$
for different maturities.
\begin{table}[ht]
\begin{center}
\begin{tabular}{|c|c|c|c|c|c|c|} \hline 
 & 6m &
1y & 2y & 3y & 4y & 5y \\
\hline
$\alpha$ & 0.00085 & 0.00165 & 0.00362 & 0.00570 & 0.00759 & 0.00970\\
  \hline
\end{tabular}
\end{center}
\vspace{2mm}
\caption{Parameters specifying the geometry of the lattice for the realized
variance process $I_t$.}
\label{t:VarSpacing}
\end{table}
It should be noted that the 
numerical complexity of the algorithm
used to obtain the
pdf of the joint process 
$(F_t,I_t)$
for different maturities is constant since 
$C$
does not change.
As mentioned earlier, it is crucial that the value 
of
$C$
be chosen large enough with respect to the spacing 
$\alpha$
so that the process cannot get to the other side of the lattice
with positive probability. 
The guiding principle has been to ensure that the probability of this
event is smaller than
$10^{-5}$.
In fact for a fixed maturity
$t$
up to five years
the formula
$$\alpha(t):=\frac{0.42^2}{C}f(t)$$
gives a definition of the lattice
spacing 
$\alpha(t)$
which has the required property
(for the graph of 
$\alpha(t)$
see Figure~\ref{fig:Alpha}).
Here the function 
$f(t)$
is the financial time introduced 
in~\cite{FXrev}
(see Figure~\ref{fig:timechange}
for the graph of
$f(t)$ 
used in calibration to the volatility surface).
The intuition behind this definition is
that the maximal value of the realized
variance increases linearly with financial
time
and is independent of the number of lattice
points used to model it.
With this choice of parameters the process 
$I_t$
wraps around with probability 
$10^{-8}$
if 
$t$
equals 6 months,
and with probability 
$10^{-5}$
when
$t$
is 5 years
(these probabilities can be made even smaller --at the expense 
of computational efficiency-- if we choose larger values
for
$C$
and smaller spacing
$\alpha(t)$).

The computational time required to obtain the joint probability
law of
$(F_t,I_t)$
for a fixed time
$t$
on a
Pentium M processor
with 1GB of RAM,
using VB.NET and relying on LAPACK for the calculation
of the spectra,
is no more than 130 seconds. The task would be to diagonalize 
201 complex matrices of the size
$152\times152$.
However the symmetry of the discrete Fourier transform
allows us to reduce the number of matrices we need to
diagonalize to 101.
Since these matrices are independent of each other,
the algorithm for obtaining the joint law is parallelizable
and the computational time can be reduced significantly 
by using multithreaded code running on several processors.

\section{Numerical results}
\label{sec:NumRes}
Let us now use the calibrated model together with 
the pricing and hedging algorithms described in~\cite{FXrev},
subsection~\ref{subsec:PricingVar} and appendix~\ref{sec:Bloc}
to perform some numerical experiments and consistency 
checks for vanilla options, forward-starts, variance swaps
and other volatility derivatives.
We conclude this section by a numerical comparison of the spectral 
method developed in Sections~\ref{sec:path} and~\ref{sec:pricing}
with a Monte Carlo pricing algorithm for volatility derivatives
(see subsection~\ref{subsec:Comp}).

\subsection{Profiles of the Greeks}
\label{subsec:Greeks}
In order to test the pricing methodology of the model for the 
underlying forward rate we pick a strip of call options with
the same notional but with
varying maturities, all struck at the current spot level.
Since the entire framework 
for the underlying is expressed in relative terms 
with respect to the current value of the index, the strike used
for this strip of options is 100. 

Because we are interested in the behaviour of our pricing algorithm
in changing market conditions, we are going to study the properties of
delta, gamma and vega, 
given by the formulae
\begin{eqnarray*}
\Delta(x) & := & 
               \frac{C_0(x+1,1)-C_0(x-1,1)}{F(x+1)-F(x-1)}, \\
\Gamma(x) & := &  
   4 \frac{C_0(x+1,1)+C_0(x-1,1)-2C_0(x,1)}{(F(x+1)-F(x-1))^2},\\
\nu(x)   &  := &  
\frac{C_0(x,0)-C_0(x,1)}{\sigma_{0}-\sigma_{1}}
\end{eqnarray*}
respectively,
as functions of the current level of spot.
Here
$x$
denotes the point in
$\Omega$
that corresponds to the spot level of the index
at time 
$0$.

This task does not pose any additional
numerical difficulties because all it 
requires is the knowledge of the probability 
distribution functions of the underlying at the relevant
maturities conditional upon the starting level, which can be any
of the points in the grid
$\Omega$. 
But if we have already priced a single option,
then these pdfs are available to us without any further 
numerical efforts.
This is because, when pricing an option, 
the algorithm described in~\cite{FXrev},
Section~3,
calculates the entire probability kernel 
for each starting point in the grid, 
even though the option pricing formula 
only requires one row of the final result.
In this situation we require all the rows of the probability kernel
so as to obtain the prices of our option,
conditional upon different spot levels, by applying the matrix
of the kernel
to the vector whose coordinates are values of the payoff calculated
at all lattice points. 
As defined by the formulae above, 
the Greeks are linear combinations of the 
coordinates of the final result of the calculation. 

\begin{figure}[hbt]
\includegraphics[width=11cm]{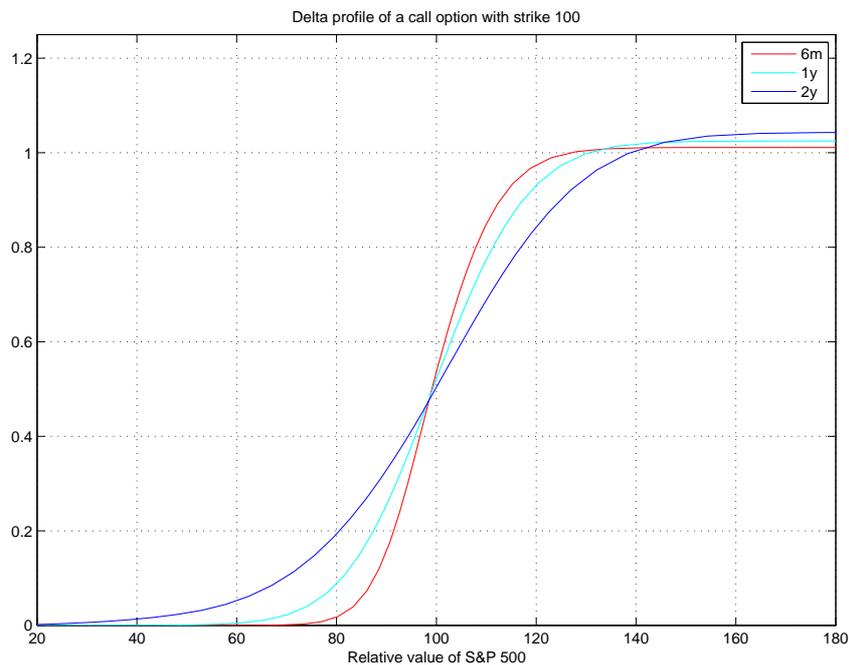}
\caption{\footnotesize{Delta profiles of call options on the 
                       S\&P 500 with maturities between
                       6 months and 2 years, all struck at 100.
                       We calculate the delta of a call
                       option (i.e.
                       $\Delta(S)=\frac{\partial C}{\partial S}(S)$)
                       for all lattice points
                       $S$
                       using a symmetric difference 
                       as described in
                       Subsection~\ref{subsec:Greeks}
                       Notice that we are using 
                       the underlying index
                       $S$  
                       and the strike of the options
                       on their relative scales with respect to the 
                       level at which the index was
                       trading when the snapshot of the market was taken.}}
\label{fig:DeltaCall}
\end{figure}

\begin{figure}[ht]
\includegraphics[width=11cm]{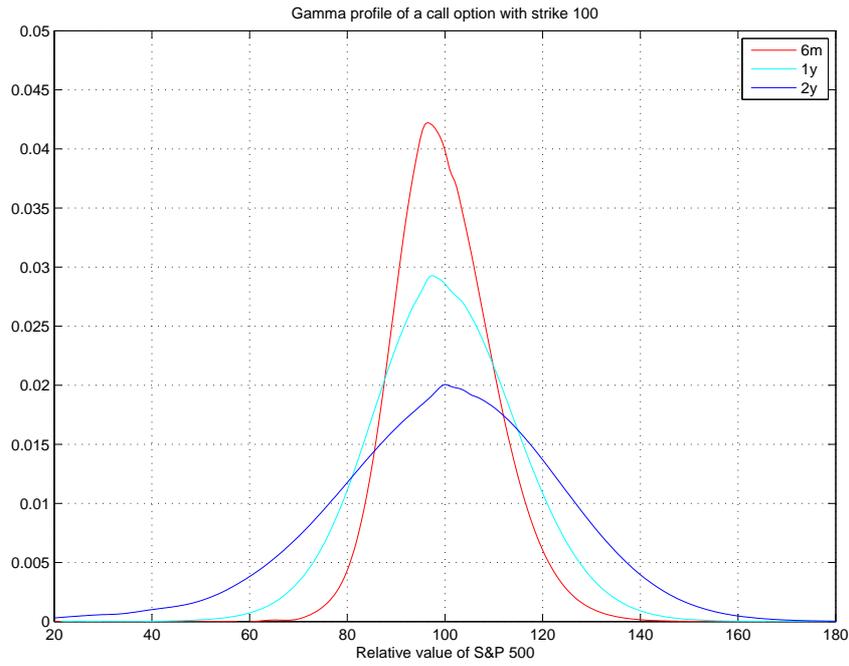}
\caption{\footnotesize{Gamma profiles of call options on the 
                       S\&P 500 with maturities between
                       6 months and 2 years, all struck at 100.
                       We calculate the gamma of a call
                       option (i.e.
                       $\Gamma(S)=\frac{\partial^2 C}{\partial S^2}(S)$)
                       for all lattice points
                       $S$
                       using a symmetric difference 
                       as described in
                       Subsection~\ref{subsec:Greeks}. 
                       The same comment as
                       in Figure~\ref{fig:DeltaCall},
                       about the relative value
                       of the index and the strike, applies.}}
\label{fig:GammaCall}
\end{figure}

\begin{figure}[hbt]
\includegraphics[width=11cm]{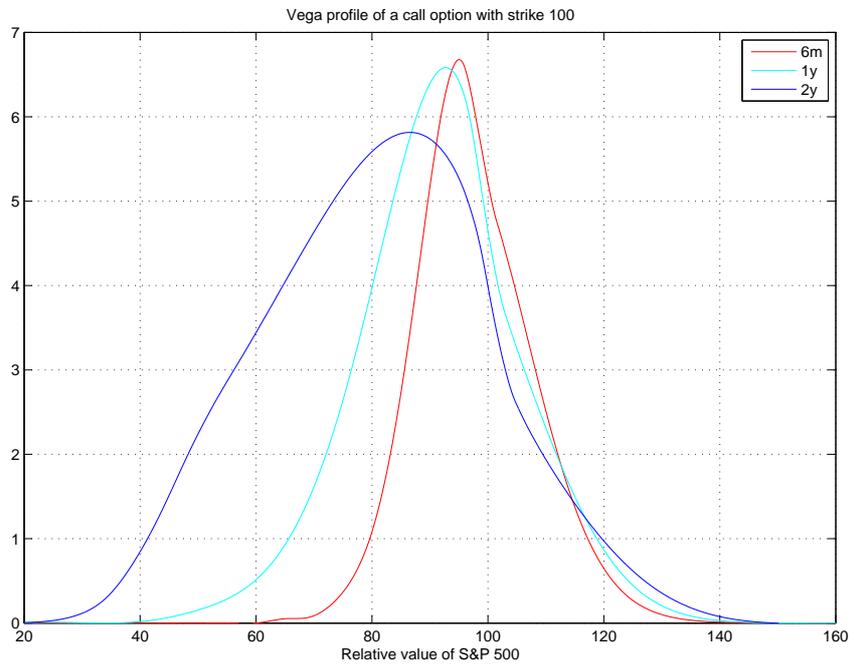}
\caption{\footnotesize{Vega profiles of call options on the 
                       S\&P 500 with maturities between
                       6 months and 2 years, all struck at 100.
                       We are calculating the vega of a call
                       option by bumping the current volatility
                       regime, repricing the option and plotting the
                       difference from the original 
                       option value,
                       for all points on the lattice.
                       Notice that vega and gamma profiles are
                       very similar in shape but different
                       in magnitude, which is consistent
                       with the general market view on the two Greeks.}}
\label{fig:VegaCall}
\end{figure}

Figure~\ref{fig:DeltaCall} contains the delta profile
of the call options for different maturities. 
Figures~\ref{fig:GammaCall} and~\ref{fig:VegaCall}
give the gamma and the vega profiles of the call options
in the strip. As expected, an owner of a 
vanilla option is both long gamma and long vega. 
The shapes
of the graphs in Figures~\ref{fig:GammaCall} 
and~\ref{fig:VegaCall} also confirm that,
according to our model,
the at-the-money options have the largest possible
vega and gamma for any given maturity.
A cursory inspection of the scales of
gamma and vega for options
with the same notional 
indicates that in our model some calendar 
spreads\footnote{A calendar spread is a structure
defined by going
long one call option and going
short another call option of the same strike
but different maturity.}
can be simultaneously long vega and short gamma
or vice versa.
 
\subsection{Forward smile} 

The forward volatility can be defined 
using the Black-Scholes formula for the
forward-starting options 
(see appendix~\ref{subsec:VIX} for details). 
One of the parameters in this formula
is a forward strike. Hence any model for the underlying process 
defines a functional relationship between forward strikes and
implied forward volatilities using the Black-Scholes pricing
formula for forward-starts, in much the same way as it defines 
the implied volatility as a function of strike. This functional
relationship is known as the \textit{forward smile}.

The reason why the forward smile is so important lies
in the fact that it determines the conditional behaviour
of the process. It is well-known that knowing all the vanilla
prices (i.e. the entire implied volatility surface)
is not enough to price path-dependent exotic options.
In terms of stochastic processes this statement can be expressed 
by saying 
that knowing the one-dimensional marginals   
of the underlying 
for all maturities does not determine the process uniquely. 
The forward smile contains the information about 
the two-dimensional distributions of the underlying process.
In other words one can have two different models
that are perfectly calibrated to the implied volatility
surface but which assign completely different values to the 
forward-starting options
(see e.g.~\cite{Schoutens1}).

Market participants can express their views on the 
two-dimensional distributions of the underlying process
by setting the prices of the forward-starting options
accordingly. It is therefore 
of utmost importance for any model used for pricing
path-dependent derivatives to have the implied forward smiles
close to the ones expected by the market.
Since the market implied forward volatilities
are rarely liquid, the main market expectation
is that the future should be similar to the present. 
Figures~\ref{fig:fs_skew3m}, \ref{fig:fs_skew6m}, \ref{fig:fs_skew1y}
and~\ref{fig:fs_skew2y}
contain the forward smiles implied by our model for maturities 
between 3 months and 2 years.
Since we have no market data to compare them with, we can only say that
the qualitative nature of the implied forward smiles is 
as expected in the following sense:
for a fixed time 
$T'$
the forward smiles are flattening with 
increasing 
$T$
(for definition of
times
$T'$
and
$T$
see appendix~\ref{subsec:VIX})
and
for a fixed difference
$T-T'$ 
the shapes of the forward smiles look 
similar when compared across maturities
$T'$.
It should be noted that the latter point 
exemplifies the stationary nature of the underlying 
model and shows that we did not have to use extreme values of
the model parameters to calibrate it to the entire 
implied volatility surface, because the two-dimensional
distributions of the underlying process have 
the stationary 
features which are expected by the market participants.

We should also note that a statistical comparison
of the forward volatility smiles implied by the model
and by the market can be carried out easily because
pricing a forward-starting option consists of two
consecutive matrix-matrix multiplications
which require little computing time.  

\subsection{Probability distribution for the implied volatility index}
One of our goals in this paper has been to describe the random
evolution of VIX through time. In Figure~\ref{fig:VIXpdf}
we plot the probability density functions of the volatility 
index, as defined at the end of appendix~\ref{subsec:VIX}.


Since we can calculate explicitly distributions of
VIX for any maturity
(like 
those in Figure~\ref{fig:VIXpdf}), 
pricing a European option 
on the VIX in our framework amounts to summing the 
values of the payoff against the pdf in the same 
way as was done 
for European payoffs on the underlying
security in~\cite{FXrev}.

Note also that the calculation of the distributions
of VIX is independent of the lifting procedure 
described in Sections~\ref{sec:path}
and~\ref{sec:pricing}.
Therefore the computational effort required to 
obtain it is minimal as the task at hand solely consists
of pricing portfolios of forward-starting options.

\subsection{Distribution of the realized variance}
\begin{figure}[ht]
\centering\noindent
       \includegraphics[width=11cm]{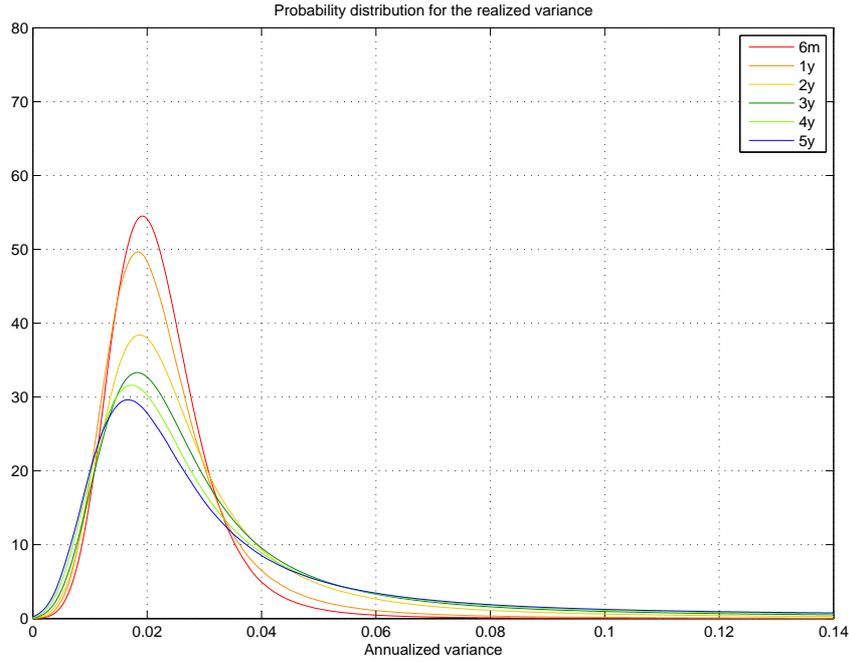}
\caption{\footnotesize{Probability distribution functions for 
                       the realized variance, quoted in annual terms,
                       for maturities between 6 months and 5 years
                       as given by our model after calibration
                       to the implied volatility surface of the 
                       S\&P 500 (see Section~\ref{sec:calibration} for 
                       details). 
                       These pdfs are marginal distributions obtained from
                       the joint probability distribution 
                       function~(\ref{eq:FinalPdf}) by integrating it in
                       the dimension of the spot value of the index.}}
\label{fig:RealizedVar}
\end{figure}

\begin{figure}[ht]
\centering\noindent
       \includegraphics[width=9cm]{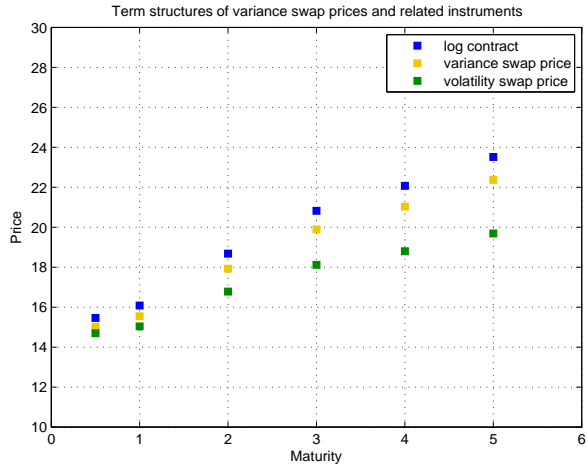}
\caption{\footnotesize{Term structures of variance 
                       swap prices 
                       for maturities between 6 months and 5 years
                       as implied by the vanilla market data using
                       the calibrated model
                       (for precise values
                       see table~\ref{tab:VarSwapData}). 
                       The delivery prices (i.e. fair
                       strikes)
                       are computed as
                       $\EE_0[\Sigma_T]$,
                       where
                       $\Sigma_T$
                       is the annualized realized variance   
                       for each tenor 
                       $T$.
                       Everything is expressed in terms 
                       of volatility, i.e. the prices are
                       in percent and are obtained by taking the 
                       square root of the variance.
                       We also plot 
                       fair delivery prices for volatility 
                       swaps.
                       The convexity bias implied by the model
                       can be clearly observed.
                       The value of the log contract, 
                       defined in equation~(\ref{eq:VarHedge}), 
                       given by the underlying model 
                       is also plotted. Observe that this portfolio
                       of options is always worth more than the corresponding
                       variance swap because, in our model, we only allow
                       for down jumps (see Section~\ref{sec:calibration}).
                       This behaviour is exactly as predicted 
                       by the analysis 
                       in~(Demeterfi et al.\hspace{-0.8mm} 1999\textit{b})
                       (equation 42) when they added a single
                       down-jump to the underlying process and studied
                       its influence on the static 
                       hedge~(\ref{eq:VarHedge}) for the variance swap.}}
\label{fig:VarSwapTerm}
\end{figure}

Most of the modelling exercise presented so far
has been directed towards finding the probability 
distribution function for the annualized 
realized variance of the underlying process.
Figure~\ref{fig:RealizedVar}
contains the pdfs for the realized variance 
for maturities between 
6 months and 5 years as implied by the model
that was calibrated to the market data for the S\&P 500. 
The corresponding term structure of the fair values of
variance swaps for these maturities is given in Figure~\ref{fig:VarSwapTerm}.
This term structure is obtained by calculating the 
expectation (in the risk-neutral measure) of the 
probability distribution functions of
Figure~\ref{fig:RealizedVar}.

Figure~\ref{fig:VarSwapTerm} also contains the current 
value of the logarithmic 
payoff, as given by~(\ref{eq:VarHedge}), for the above maturities.
In~\cite{Derman} the performance of the log payoff as a hedge instrument
for the variance swap is studied in the presence of a single
down-jump throughout the life of the contract. In equation
42 the authors show that, in this case, the log payoff is 
worth more than the variance swap.
Note that the values of the log payoff in Figure~\ref{fig:VarSwapTerm}
dominate the values of the variance swaps, as given by our model.
This agrees with~\cite{Derman} because 
our model for the underlying 
exhibits jumps down but no 
up-jumps. 
In chapter 11 of~\cite{Gatheral}
it is shown that for a compound Poisson 
process with intensity
$\lambda=0.61$
and normal jumps with 
mean
$-0.09$
and variance
$0.14$,
the difference between the current value of the logarithmic payoff
and the variance swap (both maturing in one year)
equals 
$0.3\%$
if both prices are expressed in volatility terms (i.e.
the difference between the square roots of the 
values of the annualized derivatives).
This should be compared to
$0.5355\%$
which is the corresponding difference in our
model (see table~\ref{tab:VarSwapData}).
The prices, expressed in volatility terms, seen in 
Figure~\ref{fig:VarSwapTerm}
are given by the following table:
\begin{table}[ht]
\begin{tabular}{|c|c|c|c|c|c|c|} \hline 
Maturity $T$  &   
$-\frac{2}{T}\EE_0\left[\log\left(\frac{S_T}{S_0}\right)\right]$ &
 Portfolio~(\ref{eq:VixDef}) &  $\EE_0\left[\Sigma_T\right]$ &
   $\EE_0\left[\sqrt{\Sigma_T}\right]$\\
\hline
0.5    & 15.4613  & 15.4620 & 15.0054 & 14.6960\\
1      & 16.0833  & 16.0836 & 15.5478 & 15.0378\\
2      & 18.6775  & 18.6776 & 17.9255 & 16.7808\\
3      & 20.8223  & 20.8224 & 19.8931 & 18.1161\\
4      & 22.0701  & 22.0703 & 21.0327 & 18.8013\\
5      & 23.5183  & 23.5186 & 22.3735 & 19.6923\\
  \hline
\end{tabular}
\vspace{2mm}
\caption{The random variable
$\Sigma_T$
is the annualized realized variance of
the underlying process. All the prices are 
quoted in volatility terms.
Figure~\ref{fig:VarSwapTerm}
contains the pictorial description of the data presented
in this table.}
\label{tab:VarSwapData}
\end{table}


The class of HJM-like models for volatility
derivatives
(see for example~\cite{Buehler})
require an entire term structure of variance 
swaps (like the one in Figure~\ref{fig:VarSwapTerm}) 
to be calibrated. Our approach on the other
hand implies one from the vanilla options data
(and of course some modelling hypothesis).
However,
as mentioned in Section~\ref{sec:calibration},
the parameters
$\bar\sigma_\alpha$,
for 
$\alpha=0,1$,
have a negligible effect on the model implied volatility
of vanilla strikes that are observed in the markets but have a substantial
effect on the option prices for strikes smaller than 
20.
It is clear that portfolio~(\ref{eq:VixDef}), 
and therefore the corresponding variance swap itself,
is very sensitive to these options. 
Therefore by adjusting
$\bar\sigma_\alpha$,
for 
$\alpha=0,1$,
we can calibrate to variance swaps for at least 
two maturities without distorting the implied
volatility surface of our model for strikes that 
are closer to the at-the-money strike.
In particular the steepness of the variance swap
curve implied by the model can be adjusted by
choosing different values for
$\bar\sigma_1$,
which controls the short maturities since regime 
1 is the starting regime,
and
$\bar\sigma_0$,
which has more of an influence on the far
end of the curve.

\subsection{The log contract and variance swaps in a market without jumps}
\label{subsec:NoJumps}
It is well-known that in a market 
where the underlying follows
a continuous stochastic process
the fair value of a variance swap
is equal to the value of the replicating European option with the 
logarithmic payoff given in~(\ref{eq:VarHedge})
of appendix~\ref{subsec:VIX} (see for example~\cite{Derman} or~\cite{Carr}).
In the presence of jumps this equality ceases to hold as can be observed
in Figure~\ref{fig:VarSwapTerm}. Recall from Section~\ref{sec:calibration}
that in order to calibrate our model we had to use a non-zero value for the
intensities of the down-jumps.

In this subsection the aim is to confirm that 
the prices of variance swaps and logarithmic payoffs 
agree in our framework if the randomness of the underlying model 
is based purely on stochastic and local volatility.  
To that end we set up a simplified version of the model, 
with two volatility regimes only, using the following parameters:

\begin{table}[ht]
\begin{center}
\begin{tabular}{|c|c|c|c|c|c|c|} \hline $\alpha$ & $\sigma_\alpha$ &
$\beta_\alpha$ & $ \bar\sigma_\alpha$ & $\nu^-_\alpha$ &
$\nu^+_\alpha$ & $F_\alpha$\\
\hline
0 & 10.0\%  & 0.70  & 60\%  & 0 & 0 & 100 \\
1 & 13.5\%  & 0.50  & 60\%  & 0 & 0 & 110\\
  \hline
\end{tabular}
\end{center}
\vspace{2mm}
\caption{Parameters for the local volatility regimes in the simplified
         model without jumps.}
\end{table}

\begin{figure}[ht]
\centering\noindent
       \includegraphics[width=11.5cm]{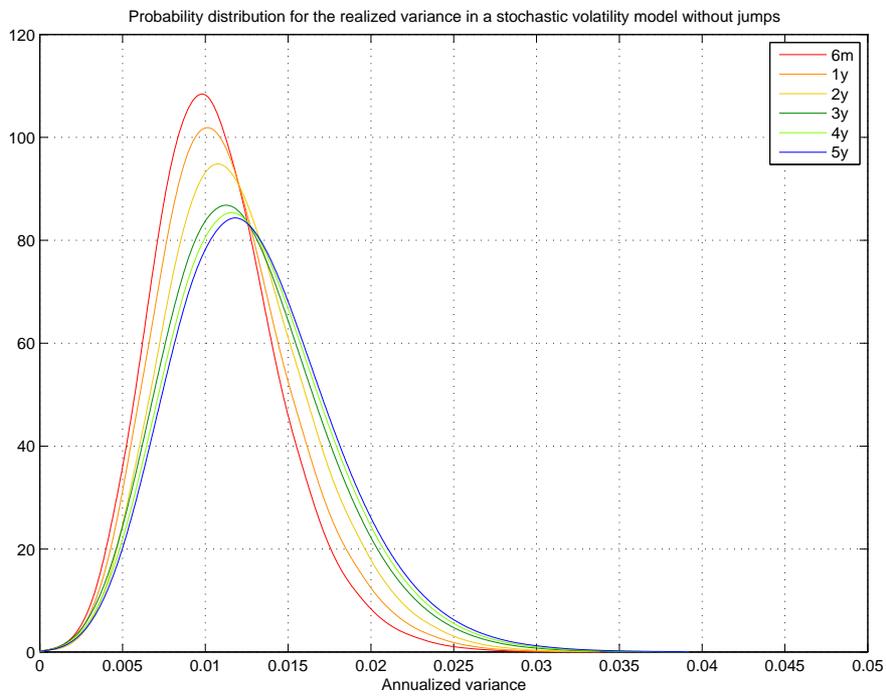}
\caption{\footnotesize{Probability distribution functions for 
                       the realized variance, quoted in annual terms,
                       for maturities between 6 months and 5 years
                       as given by the model with zero jump intensities,
                       based on two volatility regimes. For the
                       full list of values of the parameters 
                       used to specify this model see
                       Subsection~\ref{subsec:NoJumps}.}}
\label{fig:SimpleModelRealizedVar}
\end{figure}

\begin{table}[ht]
\[
\G_{\alpha}^V=
\begin{pmatrix}
-0.5    & 0.5 \\
0.5     & -0.5 
\end{pmatrix}
\]
Markov generators for stochastic volatility ($\alpha=0, 1$).
\end{table}
Note that the jump intensities in this version of the model are 
deliberately set to zero. The corresponding probability distribution
functions for maturities between 6 months and 5 years,
as implied by the simplified model, are given in 
Figure~\ref{fig:SimpleModelRealizedVar}. 

We priced variance swaps, volatility swaps, logarithmic payoffs
and structured logarithmic payoffs given  by a portfolio of vanilla
options in~(\ref{eq:VixDef}) of subsection~\ref{subsec:VIX}, 
on the notional of one dollar for maturities between 6 months and 5 years.
The results, expressed in volatility terms, can be found
in table~\ref{tab:Comp}:

\begin{table}[ht]
\begin{tabular}{|c|c|c|c|c|c|c|} \hline 
Maturity $T$  &   
$-\frac{2}{T}\EE_0\left[\log\left(\frac{S_T}{S_0}\right)\right]$ &
 Portfolio~(\ref{eq:VixDef}) &  $\EE_0\left[\Sigma_T\right]$ &
   $\EE_0\left[\sqrt{\Sigma_T}\right]$\\
\hline
0.5    & 10.3691 & 10.3753 & 10.3699 & 10.3031\\
1      & 10.6368  & 10.6450 & 10.6387 & 10.5327\\
2      & 10.9882  & 11.0003 & 10.9890 & 10.8832\\
3      & 11.2036  & 11.2142 & 11.2013 & 11.1186\\
4      & 11.3494  & 11.3614 & 11.3435 & 11.2602\\
5      & 11.4574  & 11.4691 & 11.4513 & 11.3593\\
  \hline
\end{tabular}
\vspace{2mm}
\caption{The random variable
$\Sigma_T$
is the annualized realized variance of
the underlying process.
All prices are quoted in volatility terms.}
\label{tab:Comp}
\end{table}
It follows from table~\ref{tab:Comp}
that the difference between the value of the logarithmic payoff
and the fair value of the variance swap, according to our model,
is less than 1 volatility point for all maturities. We can also observe
the quality of the approximation of the portfolio 
of options~(\ref{eq:VixDef}) to the log payoff as well as the 
convexity effect for volatility swaps. 

If the CEV parameters
$\beta_\alpha$,
for
$\alpha=0,1$,
take more extreme values, as they do in the calibrated model,
the discrepancy between the current value of the log payoff
and the fair value of the variance swap increases slightly
even if jumps have not been introduced explicitly.
We illustrate this by choosing 
the same model parameters as in 
table~\ref{t:para} of 
Section~\ref{sec:calibration}
with the exception of 
$\nu_\alpha^-$, 
for
$\alpha=0,1$,
which are taken to be 0.
Table~\ref{tab:FullModelNoJump}
shows the values of the relevant securities 
obtained using this model.

\begin{table}[ht]
\begin{tabular}{|c|c|c|c|c|c|c|} \hline 
Maturity $T$  &   
$-\frac{2}{T}\EE_0\left[\log\left(\frac{S_T}{S_0}\right)\right]$ &
 Portfolio~(\ref{eq:VixDef}) &  $\EE_0\left[\Sigma_T\right]$ &
   $\EE_0\left[\sqrt{\Sigma_T}\right]$\\
\hline
0.5    & 14.9455  & 14.9461 & 14.9437 & 14.6770\\
1      & 15.4230  & 15.4234 & 15.4181 & 15.0014\\
2      & 17.8002  & 17.8003 & 17.7871 & 16.7730\\
3      & 19.8717  & 19.8717 & 19.8510 & 18.1791\\
4      & 21.1100  & 21.1101 & 21.0827 & 18.9246\\
5      & 22.5537  & 22.5538 & 22.5171 & 19.8770\\
  \hline
\end{tabular}
\vspace{2mm}
\caption{The random variable
$\Sigma_T$
is the annualized realized variance of
the underlying process
and the full model with no jumps
(i.e. 
$\nu_\alpha^-=0$, 
for
$\alpha=0,1$).
All prices are quoted in volatility terms.}
\label{tab:FullModelNoJump}
\end{table}

\begin{figure}[ht]
\centering\noindent
       \includegraphics[width=12.5cm]{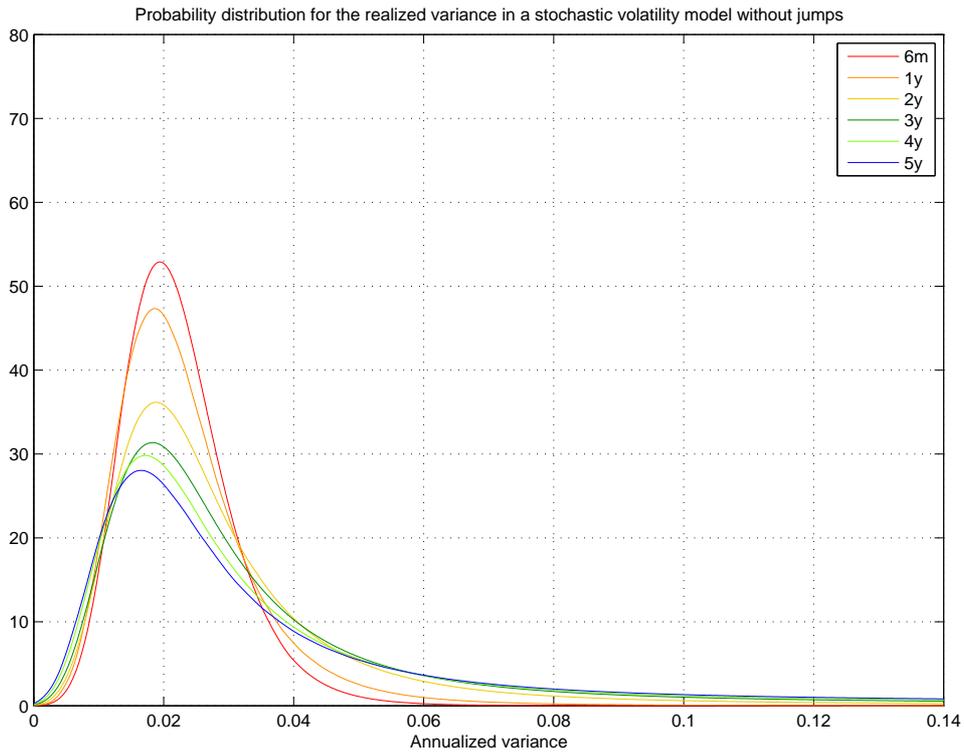}
\caption{\footnotesize{Probability distribution functions for 
                       the realized variance, quoted in annual terms,
                       for maturities between 6 months and 5 years
                       as given by the model used in 
                       Figure~\ref{fig:RealizedVar}, but  
                       with zero jump intensities,
                       based on two volatility regimes. 
                       The model parameters are the same as the 
                       ones given in table~\ref{t:para} 
                       of Section~\ref{sec:calibration}
                       but with 
                       $\nu_\alpha^-=0$, 
                       for
                       $\alpha=0,1$.}}
\label{fig:FullModelNoJumpsRealizedVar}
\end{figure}

The difference in value between the logarithmic payoff
and the variance swap, both 
maturing in 5 years,
is less than
$0.05\%$.
This should be compared with
the difference 
$0.0061\%$
which occurs
if the CEV parameters
are less extreme
(see table~\ref{tab:Comp}).
This phenomenon is due to the fact that
any random move on a lattice 
is by its definition a (perhaps small) jump. 
If local volatility function is steeper,
such moves occur more frequently thus contributing 
to the larger difference. Note however that the density functions
of the realized variance across maturities of the last example 
without jumps (see Figure~\ref{fig:FullModelNoJumpsRealizedVar})
are qualitatively very close to the densities of the realized 
variance in the calibrated model
(see Figure~\ref{fig:RealizedVar}).

\subsection{Comparison with Monte Carlo}
\label{subsec:Comp}
In this subsection we perform numerical comparisons
between the spectral method for pricing volatility
derivatives, developed in Sections~\ref{sec:path}
and~\ref{sec:pricing}, and a direct Monte Carlo method.
We generate 
$10^5$ 
paths for the underlying with a maturity of 5 years using the 
calibrated model
(for parameter values see Section~\ref{sec:calibration}),
thus obtaining an approximation for the distribution of the
realized variance for a set of maturities (1, 3 and 5 years). 
Each path consists of one point per day.
Using this
distribution we price the following derivatives:
a variance swap
$\sqrt{\EE_0\left[\Sigma_T\right]}$,
quoted in volatility terms,
a volatility swap 
$\EE_0\left[\sqrt{\Sigma_T}\right]$
and an option on the realized 
variance 
$\EE_0\left[\max\{\Sigma_T-(aK_0)^2,0\}\right]$,
where the strike
$K_0:=\sqrt{\EE_0\left[\Sigma_T\right]}$
equals the current risk-neutral mean of the realized 
variance and the paprameter 
$a$
equals 
$80\%, 100\%, 120\%$.
The results are given in the following three tables.

\begin{table}[ht]
\begin{tabular}{|l||c|c|c|c|c|c|} \hline
Maturity $T=$ 5y  &  Time (seconds) & 
$\sqrt{\EE_0\left[\Sigma_T\right]}$ & 
$\EE_0\left[\sqrt{\Sigma_T}\right]$ &
$a=80\%$ & $a=100\%$ & $a=120\%$\\
\hline 
MC: $5\cdot10^4$ paths &402&  22.48\% & 19.20\% &  2.83\% &  2.32\%  & 1.89\% \\
MC: $10^5$paths  & 780  & 22.46\% & 19.19\% & 2.81\% & 2.31\% & 1.88\%\\
Spectral method & 130  & 22.37\% & 19.69\% & 2.61\% &  2.05\% &  1.59\% \\
\hline
\end{tabular}
\end{table}

\begin{table}[ht]
\begin{tabular}{|l||c|c|c|c|c|c|} \hline
Maturity $T=$ 3y  &  Time (seconds) & 
$\sqrt{\EE_0\left[\Sigma_T\right]}$ & 
$\EE_0\left[\sqrt{\Sigma_T}\right]$ &
$a=80\%$ & $a=100\%$ & $a=120\%$\\
\hline 
MC: $5\cdot10^4$ paths & 219 & 19.95\% & 17.51\% & 2.05\%&1.61\%&1.28\% \\
MC: $10^5$ paths & 439 & 19.92\% & 17.49\% & 2.04\%&1.60\%&1.27\%\\
Spectral method & 130 & 19.89\% & 18.12\% & 1.85\%&1.34\%&0.99\% \\
\hline
\end{tabular}
\end{table}

\begin{table}[ht]
\begin{tabular}{|l||c|c|c|c|c|c|} \hline
Maturity $T=$ 1y  &  Time (seconds) & 
$\sqrt{\EE_0\left[\Sigma_T\right]}$ & 
$\EE_0\left[\sqrt{\Sigma_T}\right]$ &
$a=80\%$ & $a=100\%$ & $a=120\%$\\
\hline 
MC: $5\cdot10^4$ paths & 65 & 15.53\% & 14.16\% & 1.13\%&0.80\%&0.56\%\\
MC: $10^5$ paths & 130 & 15.53\% & 14.15\% &1.13\%&0.80\%&0.56\%\\
Spectral method      &130  &   15.55\% & 15.04\% &0.95\%&0.46\%&0.22\%\\
\hline
\end{tabular}
\end{table}

Both the spectral method and the Monte Carlo algorithm 
have been implemented in VB.NET on a Pentium M processor
with 1 GB of memory.

The spectral and Monte Carlo methods agree well on 
the prices of variance swaps because the corresponding payoff is linear
in 
the realized variance. This is to be expected since the
construction in Section~\ref{sec:path} of the continuous-time
chain
$I_t$
implies that the infinitesimal drifts of the processes 
$I_t$
and
$\Sigma_t$
coincide
(see~\eqref{eq:InstVar} and~\eqref{eq:InstDrift}), which means that
the random variables
$I_T$
and
$\Sigma_T$
are equal in expectation.
The discrepancy between the two methods in the case
of non-linear payoffs
arises because the higher infinitesimal moments 
of 
$I_t$
and
$\Sigma_t$
do not coincide. 

Note also that, as mentioned in the introduction, the spectral
method allows the calculation of the Greeks (delta, gamma,
vega) of the volatility derivatives without adding computational
complexity. In other words the gamma of the variance
swap,  for example, can be obtained for any of the maturities above in approximately
130 seconds using the same simple algorithm as 
in~\ref{subsec:Greeks}. On the other hand it 
is not immediately clear how to extend the Monte Carlo method 
to find the Greeks without substantially adding computational
complexity.

\section{Conclusion}
\label{sec:Conclusion}
In this paper we introduce an approach for pricing 
derivatives that
depend on pure realized variance (such as volatility swaps and variance swaptions)
and derivatives that are sensitive to the implied volatility smile (such
as forward-starting options)
within the same framework. 

The underlying model is a stochastic volatility
model with jumps that has the ability to switch between different CEV
regimes and therefore exhibit different
characteristics in different market scenarios.
The structure of the model allows it to be calibrated to 
the implied volatility surface with minimal explicit time-dependence.
The stationary nature of the model is best described by the implied forward
smile behaviour that can be observed in Figures~\ref{fig:fs_skew3m},
\ref{fig:fs_skew6m}, \ref{fig:fs_skew1y} and~\ref{fig:fs_skew2y}.

The model is then extended in such a way that it captures 
the realized variance of the underlying process, while retaining
complete numerical solubility. Two key ideas that make it possible
to keep track of the path information numerically are:
\begin{itemize}
\item the observation that path-dependence can be expressed as the
      lifting of a Markov generator, and that
\item the state-space of the lifted process can be chosen 
      to be on the circle so that 
      numerical tractability is retained.      
\end{itemize}
Having obtained the joint
probability distribution function for the realized variance 
and the underlying process, we outline the pricing algorithms for 
derivatives that are sensitive to the realized variance
and the implied volatility within the same model.

\appendix

\section{Diagonalization algorithm for partial-circulant matrices}
\label{sec:Bloc}
In this section our aim is to  
generalize
a known diagonalization
method from linear algebra
which will yield a numerically
efficient algorithm for obtaining 
the joint probability distribution
of the spot and realized variance
at any maturity. Let us start with some
well-known concepts.

A matrix
$C\in\RR^{n\times n}$
is \textit{circulant}
if it is 
of the form
\[
C=
\begin{pmatrix}
c_0     & c_1     & c_2    & \cdots & c_{n-1} \\
c_{n-1} & c_0     & c_1    & \cdots & c_{n-2} \\
c_{n-2} & c_{n-1} & c_0    & \ddots & \vdots \\
\vdots  & \vdots  & \ddots & \ddots & c_1 \\
c_1     & c_2     & \cdots & c_{n-1}& c_0 
\end{pmatrix},
\]
where each row is a cyclic permutation of the row above.
The structure of matrix
$C$
can also be expressed as 
$$C_{ij}=c_{(i-j)\!\!\!\!\!\mod n},$$
where 
$C_{ij}$
is the entry in the 
$i$-th
row and 
$j$-th 
column of matrix
$C$.
It is clear that any circulant matrix
is a Toeplitz\footnote{For definition
see for example~\cite{Boettcher}. Toeplitz operators
arise in many contexts in theory and practice
and therefore 
constitute one of the most important 
classes of non-self-adjoint operators. They 
provide a setting for a fruitful interplay
between operator theory, complex analysis 
and Banach algebras.} operator and in fact 
circulant matrices
are used to approximate 
general Toeplitz
matrices
and explain 
the asymptotic behaviour of their spectra.
We will not investigate this idea any further
(for more information on the topic 
see~\cite{Boettcher})
since our main interest lies in  
a different generalization of 
circulant matrices, namely that of partial-circulant matrices,
which will be defined in subsection~\ref{subsec:ParCirc}.
Before doing that we are going 
to recall some of the known 
properties of circulant matrices.

\subsection{Eigenvalues and eigenvectors of circulant matrices} 
\label{subsec:EigCirc}

Let
$C$
denote a circulant matrix of dimension
$n$
as defined above.
The eigenvalue 
$\lambda$
and
the eigenvector
$y\in\RR^n$
are the solutions of the equation
$Cy=\lambda y$,
which is equivalent to the system of 
$n$
linear difference equations with constant coefficients:
\begin{eqnarray*}
\sum_{k=o}^{n-1}c_{k}y_k & = & \lambda y_0 \>\>\>\>\mathrm{and}\\
\sum_{k=0}^{j-1}c_{n-j+k}y_k+ \sum_{k=j}^{n-1}c_{k-j}y_k & = & \lambda y_j,\>\>\>\>
\mathrm{for}\>\>\>\>j\in\{1,\ldots,n-1\}.
\end{eqnarray*}
The variables 
$y_k$,
for
$k\in\{0,\ldots,n-1\}$,
in these equations are simply the coordinates of the 
eigenvector 
$y$.
Such systems are routinely solved by guessing the 
solution and proving that it is correct (see appendix 1
in~\cite{Grimmett}). 
The solution in this case is of the form
$$\lambda=\sum_{k=0}^{n-1}c_kz^k\>\>\>\>\mathrm{and}
\>\>\>\>y_j=\frac{z^j}{\sqrt{n}}\>\>\>\>
\mathrm{for}\>\>\>\>j\in\{0,\ldots,n-1\},$$
where 
$z$
is a complex number 
which satisfies
$z^n=1$.
This implies that the eigenvalue-eigenvector pairs
of matrix
$C$
are parameterized by the 
$n$-th roots of unity
which are of the form
$z_r=\exp(-2\pi ir/n)$,
where the index 
$r$
lies in
$\{0,\ldots,n-1\}$
and 
$i$
is the imaginary unit.
Therefore the
$j$-th coordinate of the 
$r$-th eigenvector,
together with the corresponding 
eigenvalue, can be expressed as
\begin{eqnarray}
\label{eq:CirculantEig}
y^{(r)}_j & = & \frac{1}{\sqrt{n}}e^{-i\frac{2\pi}{n}rj} \>\>\>\>\mathrm{and}\\
\label{eq:CirculantEigValue}
\lambda_r & = & \sum_{k=0}^{n-1}c_k e^{-i\frac{2\pi}{n}rk}
\>\>\>\>\mathrm{for}\>\>\>\>r,j\in\{0,\ldots,n-1\}.
\end{eqnarray}

This representation is extremely useful 
because it allows us to deduce a number of fundamental
facts about circulant matrices. Let us start with the 
eigenvectors. It is obvious that if we put all 
$n$
vectors
$y^{(r)}$
side by side into a matrix,
the determinant of the linear operator
obtained is the Vandermonde determinant,
which is non-zero since its
parameters are the 
$n$
distinct solutions of the equation
$z^n=1$.
This implies that matrix 
$C$
can be diagonalized and that all its  
eigenvectors are of the form~(\ref{eq:CirculantEig}).

Another key property of circulant matrices is that they
can all be diagonalized using the same set of eigenvectors.
This follows directly from~(\ref{eq:CirculantEig})
since the expression for the vectors
$y^{(r)}$
are clearly independent of matrix
$C$.

Expression~(\ref{eq:CirculantEigValue})
tells us that 
the 
$r$-th
eigenvalue of
$C$
equals
the value (at the point
$r$)
of the discrete Fourier transform (DFT)
of the sequence
$(c_j)_{j=0,\ldots,n-1}$.
We can therefore recover the sequence
$(c_j)$
from the spectrum
$(\lambda_r)_{r=0,\ldots,n-1}$
of 
$C$
using the inverse discrete Fourier transform.
Even though this is a very well-known and celebrated fact,
we will now present a short proof for it, 
because the argument itself sheds light on the behaviour
of circulant matrices.

Note that for any index
$k\in\ZZ$,
such that 
$(k\!\!\mod n)$
is different from zero, we obtain
\begin{eqnarray}
\label{eq:Root}
\sum_{r=0}^{n-1}e^{i\frac{2\pi}{n}rk}=
\frac{1-e^{i\frac{2\pi}{n}kn}}{1-e^{i\frac{2\pi}{n}k}}=0,
\end{eqnarray}
by summing a finite geometric series.
In particular this implies that the above sum for any 
$k\in\ZZ$
equals
$n\delta_{1,k\!\!\mod n}$,
where 
$\delta$
is the Kronecker delta which takes value 
$1$
at zero
and value 
$0$
everywhere else.
The inversion formula for the DFT is now an easy consequence
\begin{eqnarray*}
\frac{1}{n}\sum_{r=0}^{n-1}\lambda_r e^{i\frac{2\pi}{n}rl} & = &
\frac{1}{n}\sum_{r=0}^{n-1}\sum_{k=0}^{n-1}
\left(c_k e^{-i\frac{2\pi}{n}rk}\right) e^{i\frac{2\pi}{n}rl}\\
& = & \sum_{k=0}^{n-1}c_k \frac{1}{n}\sum_{r=0}^{n-1}e^{i\frac{2\pi}{n}r(l-k)}
= c_l,
\end{eqnarray*}
for any
$l$
in
$\{0,\ldots,n-1\}$.
Before proceeding we should note that the argument we have 
just outlined implies that a circulant matrix
is uniquely\footnote{Such a statement is untrue
even for self-adjoint and unitary operators.} determined by its spectrum.

Another consequence of the extraordinary identity~(\ref{eq:Root})
is that for any pair of distinct indices
$k$
and
$r$
in
$\{0,\ldots,n-1\}$,
the corresponding eigenvectors
$y^{(k)}$
and
$y^{(r)}$
are perpendicular to each other.
Since we have chosen the vectors
$y^{(r)}$
in~(\ref{eq:CirculantEig}) so that 
their norm is one,
the set of all eigenvectors of a circulant
matrix
is an orthonormal
basis of the vector space
$\CC^n$.

Let
$A$
be another circulant matrix
given by the sequence
$(a_k)_{k=0,\ldots,n-1}$
with the spectrum
$(\alpha_r)_{r=0,\ldots,n-1}$. Since 
$A$
and
$C$
can be diagonalized simultaneously using 
the basis
$\{y^{(r)};\>\>r=0,\ldots,n-1\}$,
it follows that the product
$AC$
is also diagonal in this basis
and that its eigenvalues are of the 
form
$\alpha_r\lambda_r$.
Therefore 
$AC$
is a circulant matrix whose first row
equals the convolution\footnote{Recall
that the DFT of the convolution of two sequences
equals the product of the DFTs of each of the sequences.} 
of the sequences
$(a_k)$
and
$(c_k)$.
The diagonal representation also implies that the matrices
$A$
and
$C$
commute. Finally note that the sum
$A+C$
is also a circulant matrix. 

\subsection{Partial-circulant matrices}
\label{subsec:ParCirc}

We are now going to define a class of matrices,
that will include the Markov generator given by~(\ref{eq:lift}),
which can be diagonalized by the semi-analytic algorithm
from subsection~\ref{subsec:alg}.

Let
$A$
be a linear operator represented by a matrix
in
$\RR^{m\times m}$
and let 
$B^{(k)}$,
for 
$k=0,\ldots,m-1$,
be a family of 
$n$-dimensional matrices with 
the following property: there exists an invertible matrix
$U\in\CC^{n\times n}$
such that 
$$U^{-1}B^{(k)}U=\Lambda^{(k)},\>\>\>\>\mathrm{for}\>\>\mathrm{all}\>\>\>\>
k\in\{0,\ldots,m-1\},$$
where 
$\Lambda^{(k)}$
is a diagonal matrix in
$\CC^{n\times n}$.
In other words this condition stipulates
that the family of matrices
$B^{(k)}$
can be
simultaneously diagonalized by the transformation
$U$.
Therefore the columns of matrix
$U$
are eigenvectors of 
$B^{(k)}$
for all 
$k$
between 
0 
and
$m-1$.

Let us now define a large linear operator
$\widetilde{A}$,
acting on a vector space of dimension
$mn$,
in the following way. Clearly matrix 
$\widetilde{A}$
can be decomposed naturally into 
$m^2$
blocks of size 
$n\times n$.
Let 
$\widetilde{A}_{i,j}$
denote an
$n\times n$
matrix which represents 
the block in
the
$i$-th row and
$j$-th column
of this decomposition.
We now define the operator 
$\widetilde{A}$
as
\begin{eqnarray}
\label{eq:diag}
\widetilde{A}_{ii} & := & B^{(i)}+A_{ii}\II_{\RR^n}\>\>\>\>\mathrm{and}\\
\label{eq:diag1}
\widetilde{A}_{ij} & := & A_{ij}\II_{\RR^n},\>\>\>\>\mathrm{for}\>\>\mathrm{all}
\>\>\>\>i,j\in\{1,\ldots,m\}\>\>\>\>\mathrm{such}\>\>\mathrm{that}\>\>\>\>
i\neq j.
\end{eqnarray}
The real numbers
$A_{ij}$
are the entries of matrix
$A$
and 
$\II_{\RR^n}$
is the identity operator
on
$\RR^n$.
We may now state our main definition.

\begin{defin}
A matrix is termed 
\textit{partial-circulant}
if it admits a structural decomposition
as in~(\ref{eq:diag}) and~(\ref{eq:diag1})
for any matrix
$A\in\RR^{m\times m}$
and a family of 
$n$-dimensional
circulant
matrices
$B^{(k)}$,
for 
$k=0,\ldots,m-1$.
\end{defin}

The concept of a partial-circulant 
matrix is well-defined
because,
as we have seen in subsection~\ref{subsec:EigCirc},
any family of 
circulant matrices can be diagonalized by a 
unitary transformation whose columns consist 
of vectors
$y^{(r)}$,
for 
$r$
between
$0$
and
$n-1$
(see equation~(\ref{eq:CirculantEig})).

The operator 
$\widetilde{A}$
is very big indeed. The typical values that are of interest
to us for the dimensions
$m$
and
$n$
are
$150$
and
$200$
respectively.
This implies that matrix 
$\widetilde{A}$
contains 
$(150\cdot200)^2$,
i.e. around
one billion, entries.
This means that even storing 
$\widetilde{A}$
on a computer requires about
10 Gb of memory.

Our task is to find the spectrum of
the operator
$\widetilde{A}$.
Given its size and the fact that it is not 
a sparse matrix, this problem at first sight
appears not to be tractable.
But the structure of matrix
$\widetilde{A}$,
combined with the ubiquitous idea of invariant
subspaces of linear operators,
will yield the solution. We will describe the 
diagonalization algorithm for 
partial-circulant matrices 
in subsection~\ref{subsec:alg}.
Before we do this we need to recall the
basic properties of invariant
subspaces.

\subsection{Invariant subspaces of linear operators}
\label{subsec:InvSub}
Let 
$A:V\rightarrow V$
be a linear operator on a finite-dimensional vector space 
$V$.
By definition a subspace
$X$
of
$V$
is an \textit{invariant subspace}
of the operator 
$A$
if and only if 
$AX\subseteq X$.
Note that the set 
$AX$
is a subspace of 
$V$.
It is clear from the definition
that vector spaces 
$\{0\}$,
$V$,
$AV$
and 
$\mathrm{ker}(A)$
are all invariant subspaces
of the operator
$A$. Another trivial example 
is the space of all eigenvectors of 
$A$
that belong to an eigenvalue
$\lambda$.

It is the non-trivial examples however that make this concept 
so powerful. If we can find two invariant subspaces 
$X_1$
and 
$X_2$
of 
$V$
for the operator
$A$,
such that 
$X_1\cap X_2=\{0\}$
and
$\dim X_1+\dim X_2=\dim V$
(i.e. 
$V=X_1\oplus X_2$),
then in the appropriate basis
the matrix representing the operator
$A$
takes the form
\[
D=\left(
\begin{array}{cc}
A_1 & 0\\
0  & A_2
\end{array} \right),
\]
where 
$A_1$
(resp. 
$A_2$)
is the matrix 
acting on the subspace
$X_1$
(resp. 
$X_2$).
The zeros in the above expression represent trivial linear 
operators that map the subspace 
$X_1$
into the origin of the subspace
$X_2$
and vice versa.

The advantage of this structural decomposition of 
the original operator
$A$
is clear because it reduces the dimensionality 
of the problem.
The spectral decomposition (i.e. the eigenvalues and eigenvectors) of
$A$
can now be obtained from the spectral decomposition of two
smaller operators
$A_1$
and
$A_2$.
\textit{Block-diagonalization}
consists of finding the transition matrix
$F$
(i.e. the appropriate coordinate change)
that will transform the original matrix 
$A$
into block-diagonal form given by matrix
$D$
above:
$$F^{-1}AF=D.$$

\subsection{Algorithm for block-diagonalization}
\label{subsec:alg}
Let 
$\widetilde{A}$
be the linear operator defined 
in~(\ref{eq:diag}) and~(\ref{eq:diag1})
which acts on the vector space
$\CC^{mn}$.
We are now going to describe the block-diagonalization
algorithm for
$\widetilde{A}$.
In other words we are going to find 
invariant subspaces 
$V_j$
of the operator
$\widetilde{A}$
(where
$j$
ranges between 
$1$
and
$n$),
such that 
$\CC^{mn}=V_1\oplus\cdots\oplus V_n$,
and a transition matrix
$F\in\CC^{mn\times mn}$,
such that the only non-zero 
$n\times n$
blocks of matrix
$F^{-1}\widetilde{A}F$
are the diagonal ones.

Recall that, by definition of  
$\widetilde{A}$,
there exists a matrix
$U\in\CC^{n\times n}$
consisting of eigenvectors for the matrices 
$B^{(k)}$.
Put differently the columns
$u_j\in\CC^n$,
for
$j=1,\ldots,n$,
of
$U$
satisfy
the identity
$$B^{(k)}u_j=\lambda^{(k)}_j u_j\>\>\>\>\mathrm{for}\>\>\>\>k\in\{0,\ldots,m-1\}.$$
Now fix any index 
$j$
between
$1$
and 
$n$
and define vectors
$v_i^{(j)}\in\CC^{mn}$,
where
$i=1,\ldots,m$,
as follows:
\begin{eqnarray}
\label{eq:bigEig}
v_i^{(j)}:=(\underbrace{0,\ldots,0}_{(i-1)n},u_j',
\underbrace{0,\ldots,0}_{(m-i)n})',
\end{eqnarray}
where
$u_j'$
is a row of 
$n$
complex numbers obtained by transposing and conjugating
the vector
$u_j$.
We can now define the subspace
$V_j$
of 
$\CC^{mn}$
as the linear span of vectors
$v_i^{(j)}$.
It is clear that the intersection of subspaces
$V_j$
and
$V_k$
is trivial for any two distinct indices
$j,k\in\{1,\ldots,n\}$.
This is because the eigenvectors
$u_j$
and
$u_k$
are linearly independent in
$\CC^n$
since, by assumption, matrix 
$U$
is invertible. 
It follows directly from the definition that the dimension
of
$V_j$
is 
$m$.
Since there are exactly 
$n$
subspaces
$V_j$,
we obtain the decomposition
$\CC^{mn}=V_1\oplus\cdots\oplus V_n$.

If we manage to show that each space
$V_j$
is an invariant subspace for the operator
$\widetilde{A}$,
we will be able to conclude that 
$\widetilde{A}$
can be expressed in the block-diagonal form as
described in subsection~\ref{subsec:InvSub}.
Since the subspace
$V_j$
is defined as a linear span of a set of vectors
$\{v_i^{(j)};\>\>i=1,\ldots,m \}$,
the invariance property 
$\widetilde{A}V_j\subseteq V_j$
will follow if we demonstrate that the vector
$\widetilde{A}v_i^{(j)}$
is in 
$V_j$
for all 
$i=1,\ldots,m $.
By definition 
of 
$\widetilde{A}$
((\ref{eq:diag}) and~(\ref{eq:diag1}))
it immediately follows that
\begin{eqnarray}
\label{eq:BasicDiag}
\widetilde{A}v_i^{(j)}=\sum_{k=1}^mA_{ik}v_k^{(j)}+\lambda^{(i-1)}_jv_i^{(j)}
\>\>\>\>\mathrm{for}\>\>\>\>i\in\{1,\ldots,m\},
\end{eqnarray}
where 
$\lambda^{(i-1)}_j$
is the eigenvalue of matrix
$B^{(i-1)}$
that corresponds to the eigenvector
$u_j$
and the real numbers
$A_{ki}$
are the entries of matrix
$A\in\RR^{m\times m}$.
Identity~(\ref{eq:BasicDiag}) implies that each
subspace 
$V_j$
is an invariant subspace for 
$\widetilde{A}$.
Furthermore, if we define a matrix
$F\in\CC^{mn\times mn}$
in the following way
\begin{eqnarray}
\label{eq:Transition}
F:=\left(v_1^{(1)},\ldots,v_m^{(1)},v_1^{(2)},\ldots,v_m^{(2)},
\ldots,v_1^{(n)},\ldots,v_m^{(n)}\right),
\end{eqnarray}
then matrix
$D=F^{-1}\widetilde{A}F$
is block-diagonal. In other words if we decompose
$D$
into 
$n^2$
matrices 
$D_{ij}$
of size
$m\times m$,
then the following formula holds
\begin{eqnarray}
\label{eq:BlocDiagForm}
D_{ij}=\delta_{ij}(A+\Theta^{(j)})\>\>\>\>\mathrm{for}\>\>\>\>
i,j\in\{1,\ldots,n\},
\end{eqnarray}
where 
$\Theta^{(j)}$
is a diagonal matrix in
$\CC^{m\times m}$
with its
$i$-th
diagonal
element equal to
$\lambda^{(i-1)}_j$.
As usual the symbol 
$\delta_{ij}$
denotes the Kronecker delta function.

Expression~(\ref{eq:BlocDiagForm})
gives us the block-diagonal representation of 
the operator
$\widetilde{A}$.
Notice that the diagonal elements of matrix
$\Theta^{(j)}$
are precisely the eigenvalues of matrices
$B^{(i)}$,
for 
$i=0,\ldots,m-1$,
that correspond to the eigenvector
$u_j$.

The algorithm to block-diagonalize 
the operator
$\widetilde{A}$,
defined by matrices 
$A\in\RR^{m\times m}$
and
$B^{(k)}\in\RR^{n\times n}$
(see~(\ref{eq:diag}) and~(\ref{eq:diag1})),
can now be described as follows:
\vspace{3mm}
\begin{enumerate}
\item[(I)] Find matrix
           $U\in\CC^{n\times n}$
           whose columns 
           are the common eigenvectors
           $u_j$,
           for
           $j\in\{1,\ldots,n\}$,
           of the family
           $B^{(k)}$.     
\item[(II)]   Construct the transition matrix 
             $F$
             using the columns of matrix
             $U$
             as described in~(\ref{eq:bigEig})
             and~(\ref{eq:Transition}).                 
\item[(III)] Find the eigenvalues
            $\lambda^{(k)}_j$
            which satisfy
            $B^{(k)}u_j=\lambda^{(k)}_j u_j$
            for
            $k\in\{0,\ldots,m-1\}$
            and
            $j\in\{1,\ldots,n\}$.                                   
\item[(IV)] Construct diagonal matrices 
             $\Theta^{(j)}\in\CC^{m\times m}$, 
             for all           
             $j\in\{1,\ldots,n\}$,
             given by 
             $\Theta^{(j)}_{ik}=\delta_{ik}\lambda^{(i-1)}_j$,
             where the indices 
             $i,k$
             run over the set
             $\{1,\ldots,m\}$.
\item[(V)]  Construct the block-diagonal representative
             $D$
             for the operator           
             $\widetilde{A}$
             as described in~(\ref{eq:BlocDiagForm}).          
\end{enumerate}
\vspace{3mm} 

Our main task is to find the spectrum of matrix
$\widetilde{A}$.
Notice that,
since the spectrum
of
$\widetilde{A}$
is a union of the spectra of
$A+\Theta^{(j)}$,
this algorithm has reduced the problem 
of diagonalizing an
$nm\times nm$
matrix 
to finding 
the spectra of 
$n$
matrices of size
$m\times m$.
The algorithm provides a key step for our pricing
method because 
it enables us to model the behaviour of the realized
variance by increasing the numerical complexity only linearly .

We should also note that in the case of the lifted Markov
generator
$\widetilde{\L}$
in~(\ref{eq:lift}),
matrix 
$A$
is the generator 
$\L$
of the underlying process
while 
the family 
$B^{(k)}$
consists of circulant matrices.
In other words the operator
$\widetilde{\L}$
is given by a partial-circulant matrix.
It therefore follows from the discussion
in subsection~\ref{subsec:EigCirc}
that the columns of the corresponding transition matrix
$F$
are pairwise orthogonal
and that the 
entries of matrix
$\widetilde{\L} F$
are the values of a partial\footnote{Since each row of matrix
$\widetilde{\L}$
is naturally described by two variables, namely the value of the 
underlying and the value of the realized variance,
partial DFT is by definition a DFT acting on the second variable.}
discrete Fourier transform
of the rows of 
$\widetilde{\L}$.
This simple observation is useful 
when calculating the probability kernel of the 
lifted generator in Section~\ref{sec:pricing}. 

\subsection{Proof of Theorem~\ref{thm:LiftKer}}
\label{subsec:proof}

Let us start by recalling that any holomorphic function
defined on 
$\CC$
has a Taylor expansion around zero that converges everywhere.
We can therefore define 
$\phi(A)$
using the power series of the function
$\phi$
for any linear operator
$A$
on a finite-dimensional vector space.
It also follows from the fact that 
$\phi$
has a
Taylor expansion
that any invariant subspace (see 
subsection~\ref{subsec:InvSub} 
for definition)
of 
$A$
is also an invariant subspace
of 
$\phi(A)$.
In particular if 
$A$
has a block-diagonal decomposition in the sense of 
appendix~\ref{subsec:alg}, then the matrix
$\phi(A)$
also has one. Moreover
if
$B$
is a block in 
$A$,
then 
$\phi(B)$
must be a block in
$\phi(A)$.

We know that the Markov generator
$\widetilde{\L}$
can be expressed as 
$\widetilde{\L}=FDF^{-1}$,
where 
$D$
is a block-diagonal matrix of the form
\[
D=
\begin{pmatrix}
\L_0     & 0     &  \cdots & 0 \\
0        & \L_1     & \cdots & 0 \\
\vdots  & \vdots  & \ddots & \vdots \\
0     & 0     & \cdots & \L_{2C} 
\end{pmatrix},
\]
and 
the transition matrix
$F$
is given by~(\ref{eq:Transition}).
We have just seen that 
$\phi(D)$
must therefore also be in block-diagonal form:
\[
\phi(D)=
\begin{pmatrix}
\phi(\L_0)     & 0     &  \cdots & 0 \\
0        & \phi(\L_1)     & \cdots & 0 \\
\vdots  & \vdots  & \ddots & \vdots \\
0     & 0     & \cdots & \phi(\L_{2C}) 
\end{pmatrix}.
\]

The power series expansion of
$\phi$
implies
that
$\phi(\widetilde{\L})=F\phi(D)F^{-1}$.
Since matrix
$F$
is defined using the eigenvectors of circulant matrices,
it follows immediately that the inverse 
$F^{-1}$
can be obtained by transposing 
$F$
and 
conjugating each of its elements.
Note that the dimension of our circulant matrices 
is 
$2C+1$
and express
$\phi(\widetilde{\L})(x,\beta,c;y,\gamma,d)=\langle u, \phi(D)v\rangle$
as a real inner product of two vectors
$u$
and
$\phi(D)v$,
where 
$u$
equals the 
$(x,\beta,c)$-row 
of
the matrix
$F$
and
$v$
is the
$(y,\gamma,d)$-column 
of 
$F^{-1}$
(i.e. the conjugated 
$(y,\gamma,d)$-row
of
$F$).

It follows from the definition of 
$F$
and the above expression for
$\phi(D)$
that the non-zero coordinates of the vector
$u$
are of the form
$$\frac{1}{\sqrt{2C+1}}e^{-ip_kc}$$
for all 
$k\in\Psi$
and that the corresponding coordinates
of
$\phi(D)v$
are 
$$\frac{1}{\sqrt{2C+1}}e^{ip_k d}\phi(\L_k)(x,\beta;y,\gamma).$$
The equality in the theorem now follows directly from the expressions
for the coordinates of the vectors
$u$
and
$\phi(D)v$
and the definition of the real inner product.
This concludes the proof of Theorem~\ref{thm:LiftKer}.

\section{Volatility derivatives}
\label{sec:Vol}
In this appendix we are going to give a brief description of the volatility
derivatives discussed in this paper. We start with 
the simplest case, namely a forward on the realized variance,
which defines a variance swap. 
In subsection~\ref{subsec:Vol} we define options
with payoffs that are general functions of realized variance. 
Subsection~\ref{subsec:VIX} concerns derivatives
that are dependent on implied volatility. In particular
we recall the definition of the forward-starting options
and of the implied volatility index.

\subsection{Variance swaps}
\label{subsec:Var}
As mentioned above, a variance swap expiring at time
$T$
is simply a forward contract
on the realized variance
$\Sigma_T$,
quoted in annual terms,
of the underlying stock (or index)
over the time interval
$[0,T]$. 
The payoff is therefore of the form
$$(\Sigma_T-K_{\mathrm{var}})N,$$
where 
$K_{\mathrm{var}}$
is the strike and 
$N$
is the notional of the contract. The fair value of
the variance is the delivery price
$K_{\mathrm{var}}$
which makes the swap have zero value at inception.

At present such contracts are liquidly traded for most 
major indices.
The delivery price is usually quoted in the markets 
as the square of the realized volatility, i.e.
$K_{\mathrm{var}}=K^2$
where 
$K$
is a value of realized volatility expressed in percent. The notional
$N$
is usually quoted in dollars per square of the volatility 
point\footnote{A \textit{volatility point}
is one basis point of volatility, i.e. 
$0.01$
if volatility is quoted in percent. This means that the quote
for the notional
value of the variance swap tells us how much the swap owner 
gains if the realized variance increases by 
$0.0001=0.01^2$.}. 

A key part of the specification of a 
variance swap contract is how one measures
the realized variance
$\Sigma_T$.
There are a number of ways in which discretely sampled 
returns of an index (or of an 
index future\footnote{The reason for
considering index futures rather than the index itself
is twofold. The futures are used for hedging
options on the index because they are much easier
to trade than the whole portfolio of stocks 
that the index comprises. 
Also, it is well-known that futures prices are martingales
under the appropriate risk-neutral measure which depends on the
frequency of mark-to-market. If the futures contract 
marks to market continuously, then the price process
$F_t$
is a martingale in the risk-neutral measure induced by the money
market account as a numeraire. Otherwise 
we have to take the rollover strategy, with the same frequency as 
mark-to-market, as our numeraire to obtain the martingale measure
for 
$F_t$.}
$F_t$)
can be calculated and used for
defining the realized variance.
We will now describe the two most common approaches.

The usual definition of the annualized
realized (i.e. accrued) variance of the underlying process
$F_t$
in the period
$[0,T]$,
using logarithmic returns,
is 
$\frac{d}{n}\sum_{i=1}^n(\log\frac{F_{t_i}}{F_{t_{i-1}}})^2,$
where times 
$t_i$,
for 
$i=0,\ldots,n$,
are business days from
now
$t_0=0$
until expiry
$t_n=T$.
The normalization constant 
$d$
is the number of trading days per year.
Another frequently used
definition of the realized variance
is given by
$\frac{d}{n}\sum_{i=1}^n(\frac{F_{t_i}-F_{t_{i-1}}}{F_{t_{i-1}}})^2.$
It is a standard fact 
about continuous square-integrable martingales
that 
in the limit,
as we make partitions of the interval
$[0,T]$
finer and finer,
both sums exhibit the following behaviour\footnote{
Notice also that these equalities hold because the difference
of the process
$\log(F_t)$
and
$\int_0^t\frac{dF_u}{F_u}$
is of finite variation, which is a consequence of
It\^{o}'s lemma (see Theorem 3.3 in~\cite{Karatzas}).}: 
$$\langle\log F\rangle_t = \lim_{n\rightarrow \infty}
\sum_{i=1}^n\left(\log\frac{F_{t_i}}{F_{t_{i-1}}} \right)^2 =
\lim_{n\rightarrow \infty}
\sum_{i=1}^n\left(\frac{F_{t_i}-F_{t_{i-1}}}{F_{t_{i-1}}} \right)^2.
$$
The convergence here is in probability and 
the process  
$\langle\log F\rangle_t$
is the quadratic-variation\footnote{For a precise 
definition of a quadratic-variation of continuous 
square-integrable martingales see chapter 1 of~\cite{Karatzas}.} 
process
associated to
$\log(F_t)$.

In our framework the underlying process
$F_t$
is a continuous-time Markov chain (see chapter 6 of~\cite{Norris}
for definitions and basic properties). We define the annualized 
realized variance
$\Sigma_T$
(of 
$F_t$
over the time interval
$[0,T]$)
to be the limit
\begin{eqnarray}
\label{eq:DiscreteVar}
\Sigma_T:=\frac{1}{T}\lim_{\rho(n)\rightarrow 0}
\sum_{i=1}^n\left(\frac{F_{t_i}-F_{t_{i-1}}}{F_{t_{i-1}}} \right)^2,
\end{eqnarray}
where, for every 
$n\in\NN$,
the set
$(t_i)_{i=0,\ldots,n}$
is a strictly increasing sequence of times between 0 and 
$T$
and
$\rho(n):=\max\{t_i-t_{i-1};\>i=1,\ldots,n\}$
is the size of the maximal subinterval 
given by the sequence 
$(t_i)_{i=0,\ldots,n}$
(cf. Definitions~(\ref{eq:InstVar}) in Section~\ref{sec:path}
and~(\ref{eq:VarGen}), (\ref{eq:InstDrift}) in Subsection~\ref{subsec:Lifting}).
It should be noted that the techniques described in 
Sections~\ref{sec:path} and~\ref{sec:Bloc},
which provide numerical solubility for our model,
can be generalized to the situation where 
the realized variance is defined 
as a  
discretely sampled sum
in~(\ref{eq:DiscreteVar})
but
without the limit.
We are not going to pursue this line of thought any further, but 
should notice that
the discrete definition of the realized variance
would require 
the application of the block-diagonalization
algorithm (appendix~\ref{sec:Bloc}) to the probability kernel 
between any two consecutive observation times
$t_i$
rather than the 
application of the algorithm to 
the Markov generator directly, which is what 
is done in
Section~\ref{sec:pricing}.

\subsection{General payoffs of the realized variance}
\label{subsec:Vol}
A \textit{volatility swap} is a derivative given 
by the payoff 
$$(\sigma^R_T-K_{\mathrm{vol}})N,$$
where 
$\sigma^R_T$
is the realized volatility over the time interval
$[0,T]$
quoted in annual terms, 
$K_{\mathrm{vol}}$
is the annualized volatility strike
and
$N$
is the notional in dollars per volatility point. The market convention
for calculating the annualized realized volatility 
$\sigma^R_T$
differs slightly
from the usual statistical measure\footnote{Given a sample of 
$n$
values
$X_1,\ldots,X_n$
with the mean
$\mu=\frac{1}{n}\sum_{i=1}^nX_i$,
the unbiased statistical estimation of the
standard deviation is given by
$$\sqrt{\frac{1}{n-1}\sum_{i=1}^n\left(X_i-\mu\right)^2}.$$}
of a standard deviation of any discrete sample
and is given by the formula
$$\sigma^R_T=\sqrt{\frac{d}{n}\sum_{i=1}^n
\left(\frac{F_{t_i}-F_{t_{i-1}}}{F_{t_{i-1}}}\right)^2},$$
where 
$d$
is the number of trading days per year and
$t_i$
are business days from
now
$t_0=0$
until expiry
$t_n=T$
of the contract.
For our purposes we shall define realized volatility 
$\sigma^R_T$,
quoted in annual terms,
over the time interval
$[0,T]$
as
$$\sigma^R_T:=\sqrt{\Sigma_T},$$
where
$\Sigma_T$
is the annualized realized variance defined in~(\ref{eq:DiscreteVar}).
It is clear from this definition that the payoff of the 
volatility swap can be
view as a non-linear function of the realized variance.

Since volatility swaps are always entered into at equilibrium,
an important issue is the determination of the fair strike
$K_{\mathrm{vol}}$
for any given maturity
$T$.
As discussed in Section~\ref{sec:Intro}, a term structure of
such strikes must be part of the market data that some 
models require
(e.g.~\cite{Buehler}) in order to be calibrated.
In our case the strikes
$K_{\mathrm{vol}}$,
for any maturity, are implied by the model which 
uses as its calibration data 
the market implied vanilla surface.
The value of 
$K_{\mathrm{vol}}$
for a given maturity 
$T$
is then given by the expectation
$\EE_0[\sqrt{\Sigma_T}]$,
which can easily be obtained as soon as we have 
the probability distribution function
for
$\Sigma_T$.

The same reasoning applies to variance swaps.
The fair strike 
$K_{\mathrm{var}}$
for the variance swap of 
maturity 
$T$
can, within our framework, be obtained 
by taking the expectation
$\EE_0[\Sigma_T]$.
It therefore follows from the concavity 
of the square root function and Jensen's 
inequality\footnote{For any convex function
$\phi:\RR\rightarrow\RR$
and any random variable 
$X:\Omega\rightarrow\RR$
with a finite first moment
Jensen's inequality states that
$\phi(\EE[X])\leq\EE[\phi(X)]$.}
that the following relationship holds  
between the fair strikes of the variance and 
volatility swaps
$$K_{\mathrm{vol}}<\sqrt{K_{\mathrm{var}}},$$
for any maturity
$T$.
This inequality is always satisfied by the market 
quoted prices for variance and volatility swaps and 
is there to account for the fact that variance 
is a convex function of volatility. Put differently, 
this is just a convexity effect, similar
to the one observed for ordinary call options, 
related to the magnitude of 
volatility of volatility. The larger the
``vol of vol'' is, the greater the convexity 
effect becomes. This phenomenon can be observed 
clearly in the markets with a very steep skew 
for implied volatilities. If one wanted to
estimate its size in general, it would be necessary to
make assumptions about both the level and 
volatility of the future realized volatility.
Within our model this can be achieved directly 
by comparing the values of the two expectations
(see Figure~(\ref{fig:VarSwapTerm})
for this comparison based on the market implied 
vanilla surface for the S\&P 500).

There are other variance payoffs which are of practical
interest and can be
priced and hedged within our framework.
Examples are volatility and variance swaptions
whose payoffs are
$(\sqrt{\Sigma_T}-K_{\mathrm{vol}})^+$
and
$(\Sigma_T-K_{\mathrm{var}})^+$
respectively,
where as usual
$(x)^+$
equals
$\max(x,0)$
for any 
$x\in\RR$. 
Capped volatility swaps are also 
traded in the markets. Their payoff function is 
of the form
$(\min(\sqrt{\Sigma_T},\sigma_m) -K_{\mathrm{vol}})$,
where
$\sigma_m$
denotes the maximum allowed realized variance.
It is clear that all such contracts can be priced
easily within our framework by integrating any of these payoffs
against the probability distribution function (see
Figure~\ref{fig:RealizedVar}
for maturities above 6 months)
of the annualized
realized variance
$\Sigma_T$
and then multiplying the expectation
with the corresponding discount factor.

It should be noted that there exist even more exotic
products, like corridor variance swaps (see~\cite{Carr2}), 
whose payoffs depend on the variance that accrues only
if the underlying is in a predefined range.
Such products cannot be priced  directly in 
the existing framework.
A modification of the model  
would be required to deal with this class of derivatives.
However we will not pursue this avenue any further.

\subsection{Forward staring options and the volatility index}
\label{subsec:VIX}
Let 
$T'$
and
$T$
be a pair of maturities such that
$T'<T$.
A \textit{forward staring option} 
(or a \textit{forward-start}) is
a vanilla option with expiry
$T$
and the strike, set at time
$T'$,
which is equal to
$\alpha S_{T'}$.
The quantity 
$S_t$
is the underlying financial instrument 
the option is written on (usually a stock
or an index). More formally the value 
of a forward-start at time
$T$
(i.e. its payoff) is given by 
\begin{eqnarray}
\label{eq:FwdStartPayoff}
V_{FS}(T)=\left(S_T-\alpha S_{T'}\right)^+,
\end{eqnarray}
where the constant 
$\alpha$
is specified at the inception of the 
contract
and is know as the 
\textit{forward strike}. 
It is clear from the definition of the forward-start
that its value at time 
$T'$
equals the value of a plain vanilla call option
$$V_{FS}(T')=V_C(S_{T'},T-T',\alpha S_{T'})$$
that expires in 
$T-T'$
years and whose 
strike equals
$\alpha S_{T'}$.
Notice that at time
$T'$
the constant  
$\alpha$
can be characterized as the ratio\footnote{This
ratio is sometimes referred to as the ``moneyness''
of the option.}
between the spot price
$S_{T'}$
and the strike of the call option
into which the forward-start is transformed.
In the classical Black-Scholes framework, we have an 
explicit formula, denoted by
$\mathrm{BS}(S_{T'},T-T',\alpha S_{T'},r',\sigma')$, 
for the value of this call option (see~\cite{BS}).
This formula depends linearly on the spot level
$S_{T'}$
if the ratio of the spot and the 
strike (i.e. the ``moneyness'') is known.
In other words, assuming we are in the Black-Scholes 
world with a deterministic term-structure of volatility
and interest rates and zero dividends, 
we can express the value of the 
forward-start at time
$T'$
as 
$$V_{FS}(T')=S_{T'}\mathrm{BS}(1,T-T',\alpha,r',\sigma'),$$
where 
$\sigma'$
is the forward volatility rate\footnote{Assuming that
the term-structure of volatility is parametrized by
$\sigma(t)$,
the forward volatility rate 
$\sigma'$
is given by 
$\sigma'^2=\frac{1}{T-T'}\int_{T'}^T\sigma(t)^2dt.$}
and  
$r'$
is the forward interest rate between
$T'$
and
$T$.
The following key observations about the Black-Scholes value
and sensitivities of the forward-starting option are now clear:
\begin{itemize}
\item the value equals
      $V_{FS}(0)=S_{0}\mathrm{BS}(1,T-T',\alpha,r',\sigma')$
      and the delta 
      (i.e. $\frac{\partial}{\partial S_0}V_{FS}(0)$)
      is simply 
      $\mathrm{BS}(1,T-T',\alpha,r',\sigma')$,
\item the forward-starting option is gamma 
      neutral\footnote{It should be noted that, in the 
                       presence of stochastic volatility,
                       the gamma of a forward-starting option 
                       is no longer necessarily zero. However in 
                       a realistic model it should not be too 
                       large because it reflects
                       the dependence of volatility on very 
                       small moves of the underlying,
                       the effect of which should be negligible.}
      (i.e. $\frac{\partial^2}{\partial S_0^2}V_{FS}(0)=0$) 
      and 
\item the contract has non-zero vega (i.e.            
      $\frac{\partial}{\partial \sigma'}V_{FS}(0)>0$).  
\end{itemize}

\begin{figure}[ht]
        \includegraphics[width=10cm]{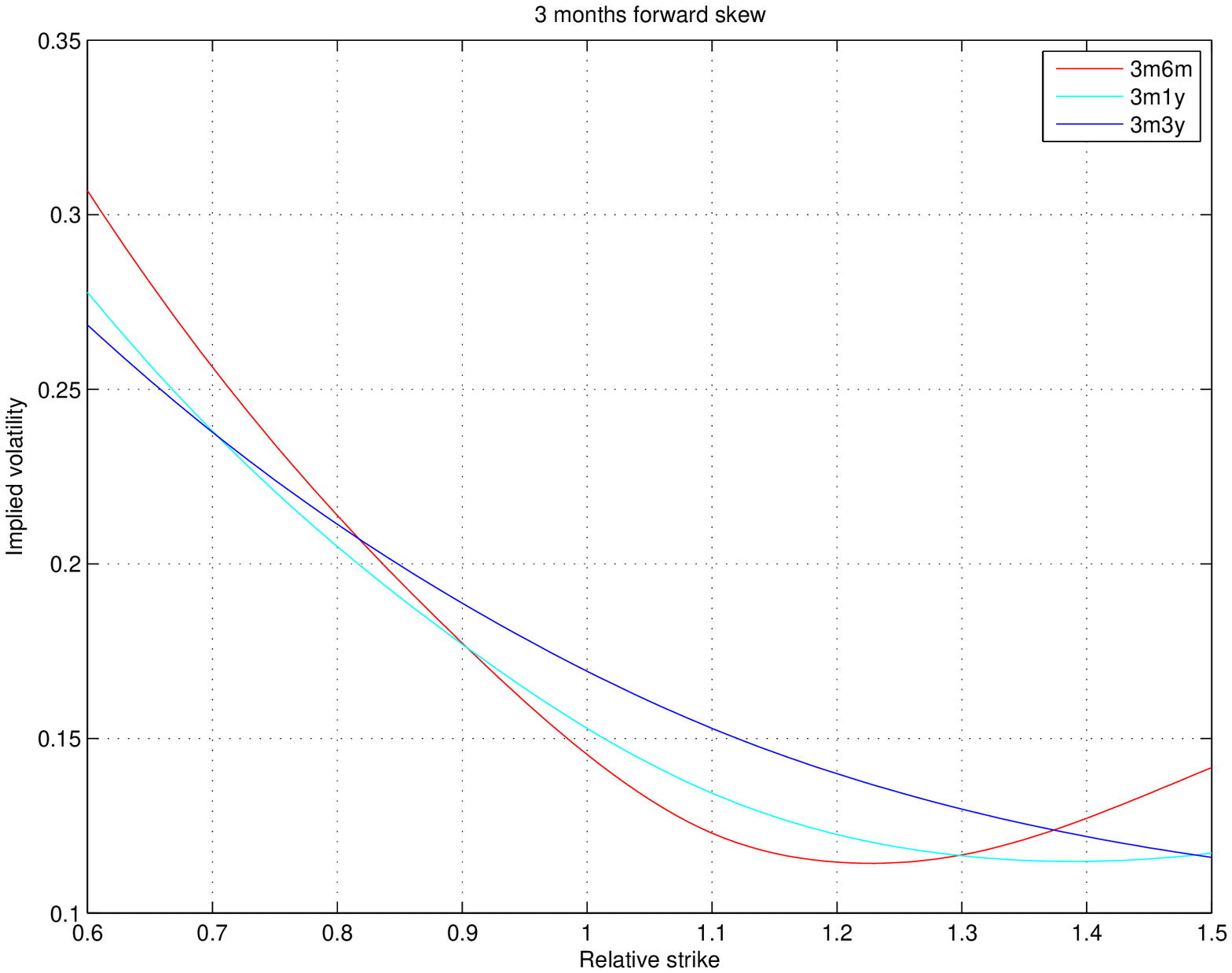}
\caption{\footnotesize{Implied forward volatility skews 
                       3 months from now.
                       For example the code 3m1y means that 
                       $T'$
                       equals 3 months and that the expiry
                       $T$
                       can be calculated as 
                       $T=(T'+1\>\mathrm{year})$,
                       where 
                       $T',T$
                       are as defined in 
                       Subsection~\ref{subsec:VIX}.   
                       Along the line of abscisse we plot
                       the forward strike 
                       $\alpha$
                       (i.e. the ``moneyness'' of the ordinary
                       call option that the forward-start becomes at
                       time
                       $T'$)
                       as defined in 
                       formula~(\ref{eq:FwdStartPayoff}).
                       The ordinate 
                       axis contains the forward
                       volatility values expressed in 
                       percentage, as impled by the model.}}
\label{fig:fs_skew3m}
\end{figure}

\begin{figure}[hbt]
       \includegraphics[width=10cm]{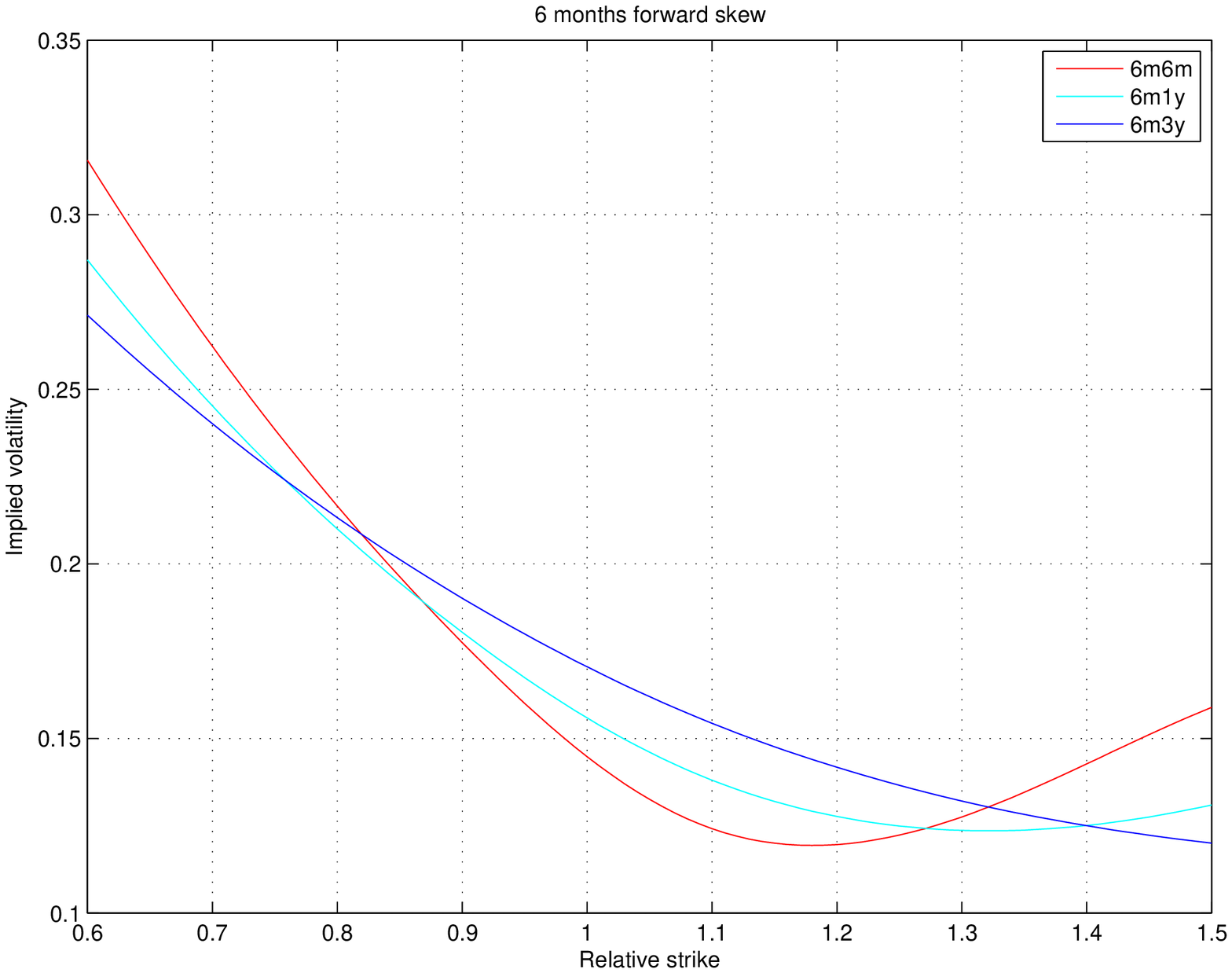}
\caption{\footnotesize{Implied forward volatility skews 
                       6 months from now.
                       For example the code 6m1y means that 
                       $T'$
                       equals 6 months and that the expiry
                       $T$
                       can be calculated as 
                       $T=(T'+1\>\mathrm{year})$,
                       where 
                       $T',T$
                       are as defined in 
                       Subsection~\ref{subsec:VIX}.   
                       Along the line of abscisse we plot
                       the forward strike 
                       $\alpha$
                       (i.e. the ``moneyness'' of the ordinary
                       call option that the forward-start becomes at
                       time
                       $T'$)
                       as defined in formula~(\ref{eq:FwdStartPayoff}).
                       The ordinate 
                       axis contains the forward
                       volatility values expressed in 
                       percentage, as impled by the model.}}       
\label{fig:fs_skew6m}
\end{figure}

\begin{figure}[hbt]
       \includegraphics[width=10cm]{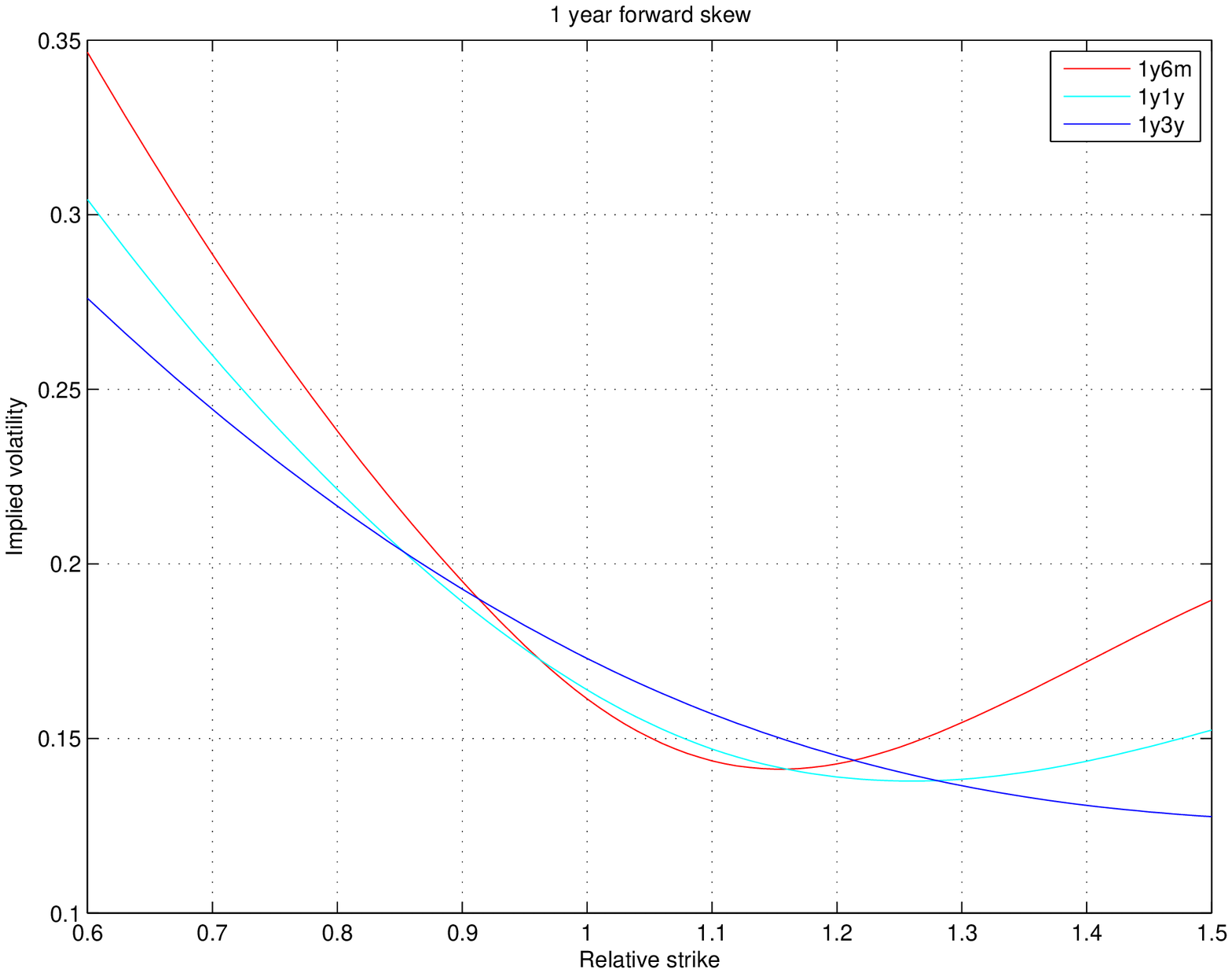}
\caption{\footnotesize{Implied forward volatility skews 
                       1 year from now.
                       For example the code 1y6m means that 
                       $T'$
                       equals 1 year and that the expiry
                       $T$
                       can be calculated as 
                       $T=(T'+6\>\mathrm{months})$,
                       where 
                       $T',T$
                       are as defined in 
                       Subsection~\ref{subsec:VIX}.   
                       Along the line of abscisse we plot
                       the forward strike 
                       $\alpha$
                       (i.e. the ``moneyness'' of the ordinary
                       call option that the forward-start becomes at
                       time
                       $T'$)
                       as defined in formula~(\ref{eq:FwdStartPayoff}).
                       The ordinate 
                       axis contains the forward
                       volatility values expressed in 
                       percentage, as impled by the model.}}
\label{fig:fs_skew1y}
\end{figure}

\begin{figure}[hbt]
\centering\noindent
       \includegraphics[width=10cm]{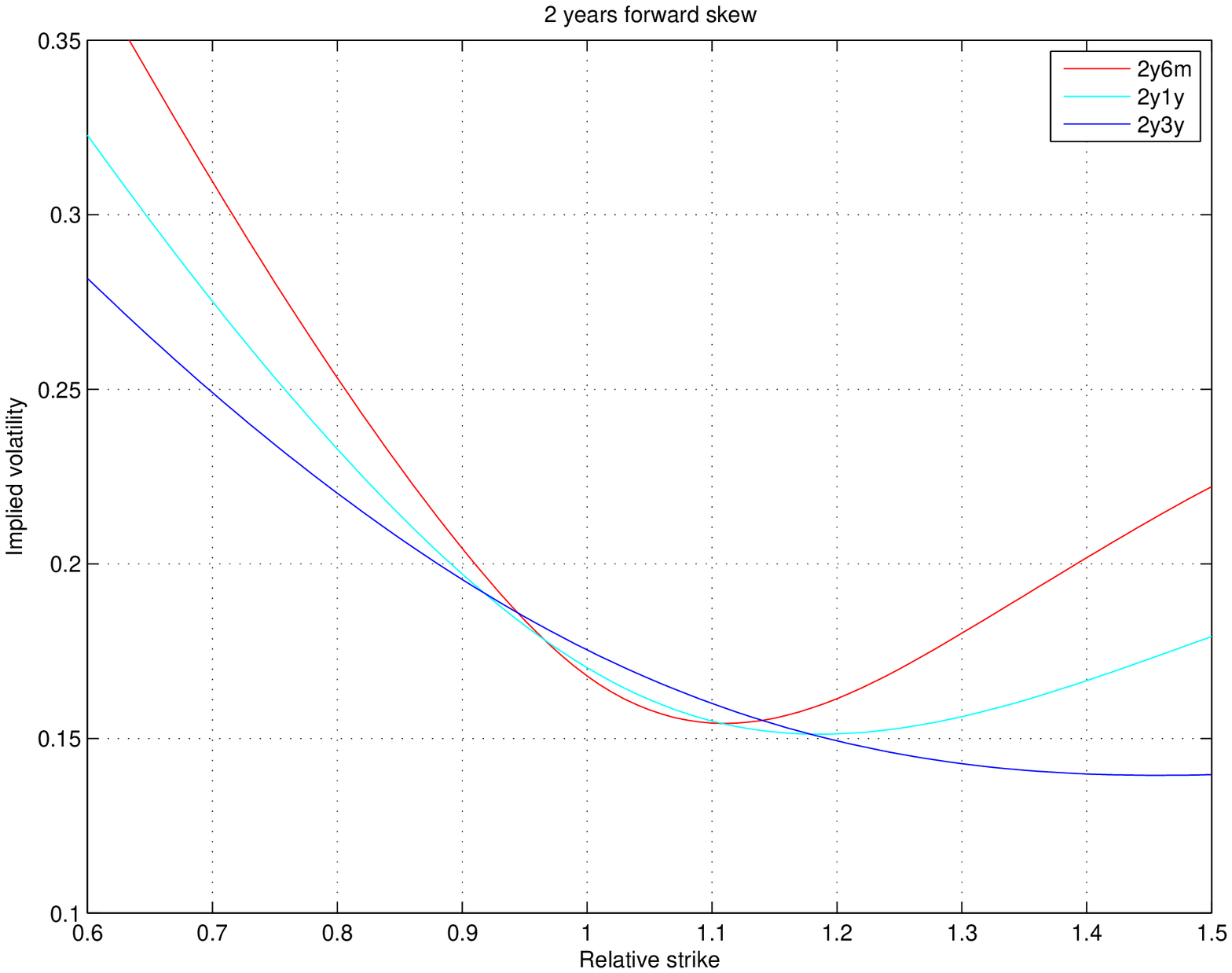}
\caption{\footnotesize{Implied forward volatility skews 
                       2 years from now.
                       For example the code 2y6m means that 
                       $T'$
                       equals 2 years and that the expiry
                       $T$
                       can be calculated as 
                       $T=(T'+6\>\mathrm{months})$,
                       where 
                       $T',T$
                       are as defined in 
                       Subsection~\ref{subsec:VIX}.   
                       Along the line of abscisse we plot
                       the forward strike 
                       $\alpha$
                       (i.e. the ``moneyness'' of the ordinary
                       call option that the forward-start becomes at
                       time
                       $T'$)
                       as defined in formula~(\ref{eq:FwdStartPayoff}).
                       The ordinate 
                       axis contains the forward
                       volatility values expressed in 
                       percentage, as impled by the model.}}
\label{fig:fs_skew2y}
\end{figure}

We are interested in the Black-Scholes pricing formula
for the forward-starts because we need to use it when expressing
the forward volatility smile of our model. 
The values of forward volatility 
$\sigma'$,
implied by the equation
$V_{FS}(0)=S_{0}\mathrm{BS}(1,T-T',\alpha,r',\sigma')$,
are plotted in 
Figures~\ref{fig:fs_skew3m},
\ref{fig:fs_skew6m}, \ref{fig:fs_skew1y} 
and~\ref{fig:fs_skew2y}
for a wide range of values of the forward strike
$\alpha$
and a variety of time horizons
$T',T$.
In other words, we first calculate the value 
$V_{FS}(0)$
of the forward-start and then 
invert the Black-Scholes pricing formula 
to obtain the implied forward volatility 
$\sigma'$.

We shall now give a brief description of 
the implied volatility index (VIX)
and then move on 
to discuss the future probability distribution of VIX,
which, as will be seen, is directly related to forward-starting 
options.

VIX was originally introduced in 1993 by Chicago Board Options Exchange 
(CBOE) as an index reflecting the 1 month implied volatility of the 
at-the-money
put and call options
on S\&P 100. 
To facilitate trading in VIX, 
in 2003 CBOE introduced a new calculation, using 
a full range of strikes for
out-of-the-money options, to define the value of VIX. At the same time
the underlying financial instrument on which the options are written 
was changed to S\&P 500
(for a detailed description of these changes and their ramifications
see~\cite{VIX}).
The new formula is 
\begin{eqnarray}
\label{eq:VixDef}
\sigma_{\mathrm{VIX}}^2=\frac{2}{T}\sum_i\frac{\Delta K_i}{K_i^2}e^{rT}Q(K_i)
-\frac{1}{T}\left(\frac{F}{K_0}-1\right)^2,
\end{eqnarray}
where the index itself is given by
$\mathrm{VIX}=100\sigma_{\mathrm{VIX}}$.
The sequence
$K_i$
consists of all the exchange quoted strikes and the quantities 
$Q(K_i)$
are the corresponding values of out-of-the-money 
put/call options expiring at maturity
$T$,
where 
$T$
equals 1 
month
The quantity 
$F$
in formula~(\ref{eq:VixDef}) is the forward value 
of the S\&P 500 index derived 
from option prices\footnote{See page 3 in~\cite{VIX} for the precise definition.}
and the at-the-money strike
$K_0$
is defined as the largest strike below
$F$.
Note also that it is precisely at 
$K_0$
that 
the symbol
$Q(K)$
in~(\ref{eq:VixDef})
changes from put to call options. 

The reason why formula~(\ref{eq:VixDef})
allows easier trading of volatility follows 
from the simple observation that 
$\sigma_{\mathrm{VIX}}^2$
is
essentially the value of a European derivative,
expiring at time
$T$,
with the 
logarithmic payoff given in~(\ref{eq:LogDecom}).
This is a consequence of
the well-known decomposition of any twice
differentiable 
payoff described in~\cite{Breeden}, \cite{Carr}, \cite{Derman}
and other sources:
\begin{eqnarray}
\label{eq:LogDecom}
-\log\left(\frac{S_T}{F}\right)=-\frac{S_T-F}{F}+\int_0^F\frac{1}{K^2}(K-S_T)^+dK
+\int_F^\infty\frac{1}{K^2}(S_T-K)^+dK.
\end{eqnarray}
This formula holds for any value of 
$F$,
but expressions simplify if we assume that 
$F$
equals 
the forward of the index 
$S_t$
at time 
$T$
(i.e.
$F=\EE[S_T]$).
By taking the expectation with respect to the risk-neutral measure
we get the following expression for the forward price of the log payoff:
\begin{eqnarray}
\label{eq:VarHedge}
-\EE\left[\log\left(\frac{S_T}{F}\right)\right]
=\int_0^F\frac{1}{K^2}e^{rT}P(K)dK
+\int_F^\infty\frac{1}{K^2}e^{rT}C(K)dK,
\end{eqnarray}
where 
$C(K)=e^{-rT}\EE[(S_T-K)^+]$
(resp.
$P(K)=e^{-rT}\EE[(K-S_T)^+]$)
is the price of a call (resp. put) option struck at 
$K$.
It is shown  
in~\cite{Derman}
that the portfolio of vanilla options given by~(\ref{eq:VarHedge})
can be used to hedge perfectly a variance swap 
if there are no jumps in the underlying market.
From our point of view the expression~(\ref{eq:VarHedge})
is interesting because a simple calculation shows that 
definition~(\ref{eq:VixDef})
is a possible discretization of it. 
By defining its own version of the
approximation to the logarithm,
CBOE has created a volatility index which can be replicated by trading
a relatively simple European payoff. This feature greatly simplifies 
the  trading of VIX. 

\begin{figure}[hbt]
\centering\noindent
       \includegraphics[width=9cm]{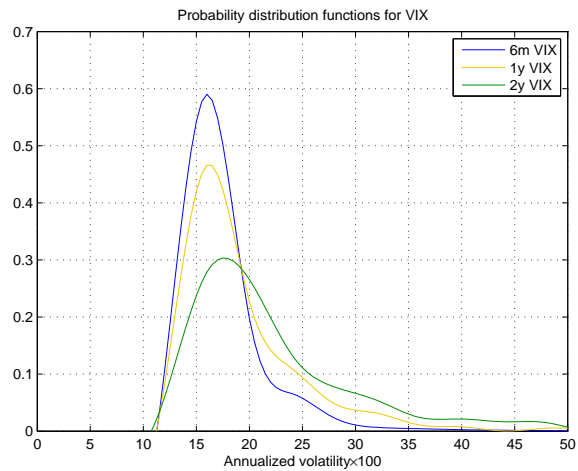}
\caption{\footnotesize{Probability distribution functions of the
                       implied volatility index for maturities 
                       between 6 months and 2 years.}}
\label{fig:VIXpdf}
\end{figure}

Since the implied volatility index
is defined by a portfolio of puts and calls in~(\ref{eq:VixDef}), 
it is clear that the random nature of the value of
VIX 
at time
$t$
will be determined by
the value of the corresponding portfolio of
forward-starting options. 
The probability distribution
function for the behaviour of the volatility index at
$t$
is obtained from the model by the following procedure:
\begin{enumerate}
\item[(I)]   Fix a level $S$ of the underlying.
\item[(II)]  Find the probability 
             that the price 
             process is at level
             $S$ at a given time 
             $t$
             in the future.
\item[(III)] Evaluate the portfolio of options 
             that define the volatility index 
             between times 
             $t$
             and
             $t+T$,
             conditional on the process 
             $S_t$
             being at level
             $S$.                
\item[(IV)]  Repeat these steps for all attainable
             levels
             $S$
             for the underlying Markov chain
             $S_t$.
\item[(V)]   Subdivide the real line into intervals with disjoint
             interiors of length
             $\delta$,
             where 
             $\delta$
             is a small positive number.
             To each of the intervals assign a probability 
             that is a sum of probabilities in step (II)
             corresponding to 
             the values obtained in step (III) that lie
             within the interval.          
\end{enumerate}
\vspace{1.5mm}
This describes the construction of the probability distribution
function of the volatility index at time 
$t$.
The plot of this pdf for a variety 
of maturities can be seen in Figure~\ref{fig:VIXpdf}.

Our final task is to price any European payoff written on
the level of VIX at a certain time horizon. 
Given that our model allows us to extract
the pdf of VIX for any
expiry, pricing such a derivative amounts to integrating the 
payoff function against the probability distribution that was
described above.


%

\bibliographystyle{amsplain}
\bibliography{cite}
\nocite{*}

\clearpage

\begin{figure}[hbt]
        \includegraphics[width=11cm]{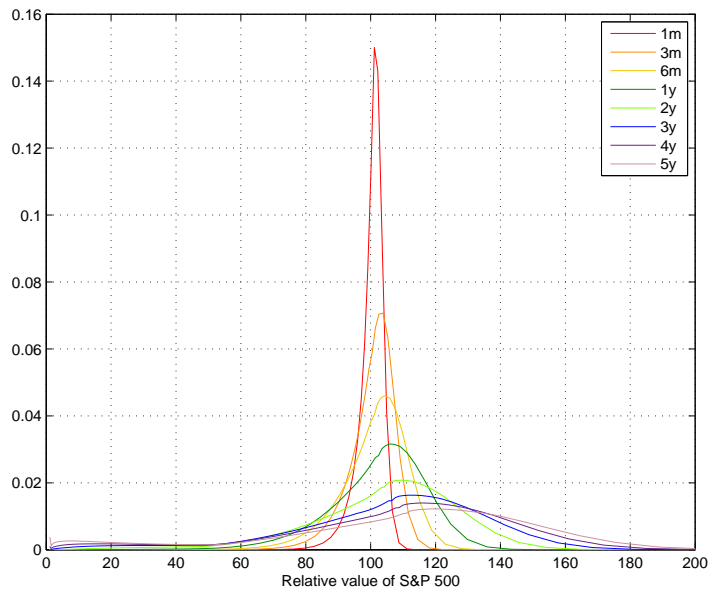}
\caption{\footnotesize{Implied probability density function 
                       for the S\&P 500 under 
                       the forward measure. The value of the S\&P 500 
                       plotted along the
                       line of abscisse is a 
                       relative value (in percent) with 
                       respect to the current level of spot.
                       The pdfs between 6 months and 5 years can also
                       be viewed as the rescaled marginals of the joint pdfs
                       in figures~\ref{fig:Joint_6m}, 
                       \ref{fig:Joint_2y}, 
                       and~\ref{fig:Joint_4y}.
                       }}
\label{fig:pdf}
\end{figure}

\clearpage

\clearpage

\clearpage

\clearpage

\clearpage

\vspace{0.7cm}

\clearpage

\clearpage

%
%

%
%
%
%
%

%

\clearpage

\clearpage

\begin{figure}[hbt]
\centering\noindent
       \includegraphics[width=9cm]{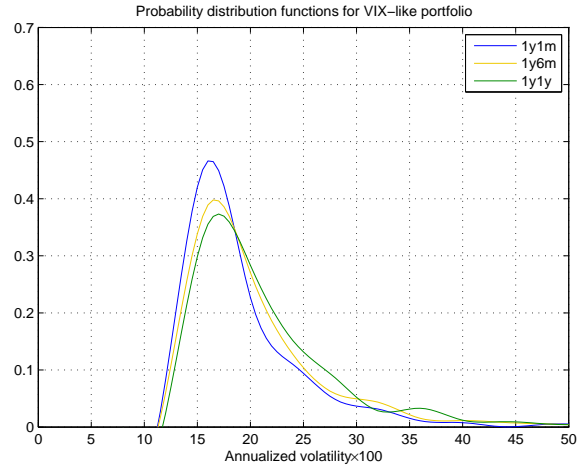}
\caption{\footnotesize{Probability distribution functions of
                       a portfolio of forward-starting options
                       as in the definition of VIX (see~(\ref{eq:VixDef})
                       in Subsection~\ref{subsec:VIX}),
                       where time 
                       $t$
                       is fixed at 1 year
                       and
                       time 
                       $T$
                       varies form 1 month to 2 years
                       (see Subsection~\ref{subsec:VIX}
                       for the definition of a probability 
                       distribution function for a portfolio of 
                       forward-starts and the role of parameters
                       $t$
                       and
                       $T$).
                       Note that the 1y1m pdf is, according to our
                      definition, the distribution of
                      VIX in 1 year.}}
\label{fig:VIXLikepdf}
\end{figure}

\end{document}